\documentclass[aps,amsmath,amssymb,twocolumn,floatfix]{revtex4-2}

\usepackage{graphicx}% Include figure files
\usepackage{dcolumn}% Align table columns on decimal point
\usepackage{bm}% bold math
\begin{document}

\title{Electronic and magnetic properties of LaCu$_{x}$Sb$_{2}$ tuned by Cu occupancy} % Force line breaks with \\

%\thanks{A footnote to the article title}%

\author{Suyoung Kim, Obinna P. Uzoh, and Eundeok Mun}

%\email{emun@sfu.ca}

\affiliation{Department of Physics, Simon Fraser University, 8888 University Drive, Burnaby, B.C. Canada}

%\date{\today}

\begin{abstract}
{We report thermodynamic and transport properties of LaCu$_{x}$Sb$_{2}$ ($0.92 \leq x \leq 1.12$), synthesized by controlling the initial loading composition and investigated by magnetization, electrical resistivity, and specific heat measurements. The physical properties of this system are highly dependent on Cu-site occupancy $x$, where residual resistivity ratio (RRR), magnetoresistance (MR), superconducting transition temperature ($T_{c}$), and electronic specific heat coefficient ($\gamma$) indicate a systematic variation as a function of $x$. The Shubnikov-de Haas quantum oscillations are observed in magnetoresistance measurements for samples close to the Cu stoichiometry $x \sim 1$, while the de Haas-van Alphen oscillations are detected in a wide range of $x$ ($0.92 \leq x \le 1.12$). For $H \parallel c$, the oscillation frequency indicates a clear $x$-dependence, implying a systematic change of Fermi surface. DFT calculations for the sample closest to ideal Cu stoichiometry reveal electronic structures with a common feature of the square-net-based semimetals, which is in good agreement with the experimental observations. The magnetic response of LaCu$_{x}$Sb$_{2}$ to magnetic fields is anisotropic owing to the Fermi surface anisotropy. Our results show how the physical properties are influenced by the Cu-site occupancy $x$, linked to the electronic bands arising from the Sb square net.}
\end{abstract}

%\keywords{Suggested keywords}%Use showkeys class option if keyword
                              %display desired
\maketitle

\section{Introduction}

The family of layered intermetallics with the chemical formula $AT$Sb$_2$ ($A =$ alkaline-earth and rare-earth, $T =$ transition metal) have attracted a significant interest due to their emergent properties and potential connection to unconventional topological superconductivity, accompanied by the electronic structure of the Sb square nets~\cite{Square-Net-1, Square-Net-2, SC and Dirac}. In this family, topologically non-trivial band structures lead to the observation of a large linear magnetoresistance \cite{Wang2012_SrZnSb2, Wang2020_EuZnSb2, LaAgSb2-multiband, ref1}, very small effective mass \cite{Liu2019_SrZnSb2, Wang2018_YbMnSb2, Liu2017_BaMnSb2, He2017_CaMnSb2, Liu2017_SrMnSb2, Huang2017_BaMnSb2, ref1}, high carrier mobility \cite{LaAgSb2-multiband, Wang2018_YbMnSb2, Liu2017_BaMnSb2, He2017_CaMnSb2, Liu2017_SrMnSb2, Huang2017_BaMnSb2}, and non-trivial Berry phase \cite{Liu2019_SrZnSb2, Wang2018_YbMnSb2, Liu2017_BaMnSb2, He2017_CaMnSb2, Liu2017_SrMnSb2, Huang2017_BaMnSb2}. In addition, ground states for moment bearing compounds appear to be antiferromagnetic (AFM)~\cite{Wilde2022_EuMnSb2, Wang2020_EuZnSb2, Liu2017_BaMnSb2, He2017_CaMnSb2}, ferromagnetic (FM)~\cite{Wang2018_YbMnSb2}, or a combination of both~\cite{Liu2017_SrMnSb2, Huang2017_BaMnSb2}. In particular, a charge density wave (CDW)~\cite{LaAgSb2-Myers1999, LaAgSb2-C.song2003, LaAgSb2-C.S.Lue2007, LaAgSb2-R.Y.Chen2017, LaAuSb2-2019, LaAuSb2-2020, LaAuSb2-2021, LaAgSb2_osillation2, LaAuSb2_osillation} and superconductivity (SC)~\cite{LaAgSb2.K.Akia, LaAuSb2-Tc, ref2, ref3, ref7} have been observed in non-magnetic La$T$Sb$_2$ ($T$ = Cu, Ag, Au) compounds.

In LaAgSb$_{2}$ compound two CDW transitions occur around $T_{CDW}^{1} = 210$~K and $T_{CDW}^{2} = 186$~K, with a superconducting phase arising at $0.3$~K in ambient pressure \cite{LaAgSb2.K.Akia, LaAgSb2-Myers1999, LaAgSb2-C.S.Lue2007, LaAgSb2-C.song2003, LaAgSb2-R.Y.Chen2017}. The CDW transition can be tuned by Au or Cu substitutions \cite{AuCusub-LaAgSb2, LaAg1-xAuxSb2-2022}. Under external pressure, a superconducting dome appears with a maximum $T_{c} \sim 1$~K at which the CDW transition suppresses \cite{LaAgSb2.K.Akia}. It has been suggested that the superconductivity under pressure is related to the non-trivial band topology~\cite{LaAgSb2.K.Akia}, hosting Dirac Fermions~\cite{LaAgSb2 de Hasas-van Alphen, LaAgSb2-multiband, LaAgSb2-2022}. At ambient pressure, the LaAuSb$_{2}$ compound shows two CDW transitions at $T_{CDW}^{1}$ = 120 K and $T_{CDW}^{2}$ = 80 K ~\cite{LaAuSb2-2019, LaAuSb2-2020, LaAuSb2-2021} and a superconducting transition at $T_{c} \sim 0.6$~K~\cite{LaAuSb2-Tc}. Under hydrostatic pressure, $T_{c}$ in LaAuSb$_{2}$ increases to $\sim$1.05~K at which the CDW transition suppresses \cite{LaAuSb2-Tc}. Interestingly, LaCuSb$_{2}$ indicates no CDW transitions, however shows the highest $T_{c} = 1$~K among the three compounds. Such a variation of CDW and SC in La$T$Sb$_{2}$ ($T$ = Ag, Au, Cu) family can be related to their unit cell volume, where the volume decreases from 209.21 \AA$^3$~\cite{LaAg1-xAuxSb2-2022} to 205.28 \AA$^3$~\cite{LaAuSb2-1996} and 196.12 \AA$^3$~\cite{ref1} as the compound is traversesd from $T$ = Ag to $T$ = Au and Cu, respectively. In addition, the formation of CDW in this family has been connected to their Fermi surface dimensionality, where the obtained Fermi surfaces from electronic structure calculations are shown to be two dimensional for $T$ = Ag and three dimensional for $T$ = Cu~\cite{LaTSb2-2014}. Furthermore, LaAgSb$_{2}$ has a full occupancy on the Ag-site, whereas LaAuSb$_{2}$ shows a Au-site deficiency, leading to significant variations in both lattice parameters and CDW transition temperatures \cite{LaAuSb2-2020}. It has also been reported in LaCu$_{x}$Sb$_{2}$ that the off-stoichiometry ($x$) on the Cu-site influences the ground states and electronic structures~\cite{ref3}.

%Table I
\begin{table*}[ht]
\caption{Summary of initial loading compositions and parameters obtained from EDX and XRD}
\begin{tabular}{l l c c c c c l}
\hline
\hline
No.~~~ & ~~~loading composition~~~ & \multicolumn{2}{c}{Lattice Parameter}  & \multicolumn{3}{c}{EDX} & ~~~LaCu$_{x}$Sb$_{2}$~~~~~ \\ \cline{3-8}

        &                  & $a$ (\AA) ~~& $c$ (\AA)~~ & ~~~La (\%)~~~ & ~~~Cu (\%)~~~ &~Sb (\%)~~~ & $x$ EDX ~~(MR) \\
\hline
1      &   La$_{0.03}$Cu$_{0.03}$Sb$_{0.91}$   			 &  4.3789(3)  &  10.2436(2)   & 19.90    & 23.40  & 56.60	& 0.92 \\
2      &   La$_{0.05}$(Cu$_{0.06}$Sb$_{0.94}$)$_{0.95}$   &  4.3761(4)  &  10.2783(4)   & 19.70    & 24.00  & 56.30 	& 0.95  \\
3      &   La$_{0.05}$(Cu$_{0.1}$Sb$_{0.9}$)$_{0.95}$       & 4.3683(8)   &  10.3485(8)   & 19.66    & 24.48  & 55.86 	& 0.97  \\
4      &   La$_{0.05}$(Cu$_{0.2}$Sb$_{0.8}$)$_{0.95}$       &  4.3680(2)  &  10.3741(9)   & 19.50    & 25.10  & 55.40 	& 1.01   \\
4-1   &   La$_{0.08}$(Cu$_{0.2}$Sb$_{0.8}$)$_{0.92}$	 &  4.3665(8)  &  10.3735(6)   & 19.50    & 25.00  & 55.50 	& 1.00~~~~ (1.005) \\
4-2   &   La$_{0.12}$(Cu$_{0.2}$Sb$_{0.8}$)$_{0.88}$   	 & 4.3657(4)   &  10.3717(4)   & 19.50    & 25.20  & 55.30 	& 1.01~~~~~(0.987)  \\
5      &   La$_{0.05}$(Cu$_{0.3}$Sb$_{0.7}$)$_{0.95}$     	 &  4.3723(9)  &  10.3820(7)   & 19.20    & 26.10  & 54.70 	& 1.06   \\
6      &   La$_{0.05}$(Cu$_{0.4}$Sb$_{0.6}$)$_{0.95}$     	 &  4.3733(5)  &  10.3870(6)   & 19.00    & 26.30  & 54.60 	& 1.07   \\
7      &   La$_{0.05}$(Cu$_{0.5}$Sb$_{0.5}$)$_{0.95}$      	 &  4.3781(3)  &  10.3958(9)   & 18.60    & 27.20  & 54.20 	& 1.12   \\ 
\hline
\hline
\end{tabular}
\label{Table1}
\end{table*}

Although the CDW state is absent in LaCuSb$_{2}$, several studies have clearly observed a conventional phonon mediated superconductivity below 1~K~\cite{ref2, ref3, ref7}. In contrast, another study detected no superconductivity down to temperatures far below 1~K, instead observed a large linear magnetoresistance (MR) in low fields and suggested a possible formation of Dirac state~\cite{ref1}. Recently, it was shown from LaCu$_{x}$Sb$_{2}$ compounds that the supercouducting state forms a dome-like shape as a function of Cu-occupancy $x$, serving as a proper candidate for studying the interplay between Dirac state and unconventional SC~\cite{ref3}. Thus, it is not unreasonable to assume that the observed inconsistent results~\cite{ref1,ref2,ref3,ref7} can be related to the Cu-site occupancy $x$. In this study, we use a combination of thermodynamic and transport property measurements of LaCu$_{x}$Sb$_{2}$ and density functional theory (DFT) to resolve the inconclusive observations, regarding the presence of superconductivity and the nature of the Dirac states in LaCu$_{x}$Sb$_{2}$. Our results clearly show that the physical properties of LaCu$_{x}$Sb$_{2}$ vary systematically as a function of $x$. Moreover, quantum oscillations detected in magnetoresistance and magnetization isotherms are consistent with DFT calculations.

\section{Experimental}

\begin {figure*}
\includegraphics[width=1\linewidth]{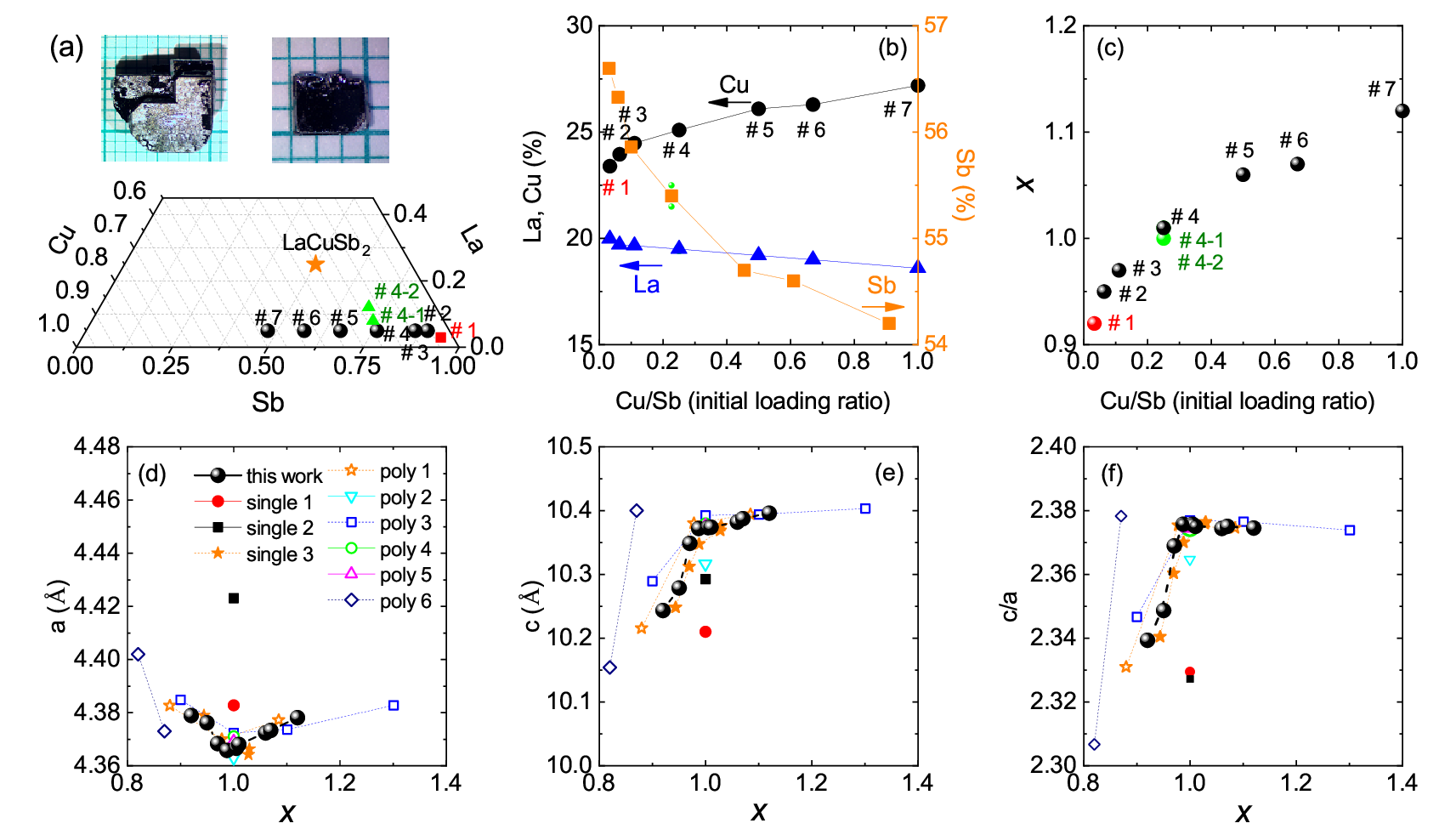}
\caption {(a) La-Cu-Sb ternary phase diagram. Symbols with numbers represents the initial loading composition in the growth. Top shows the photographs of as-grown crystals on mm grid. (b) Chemical compositions (\%) of grown single crystals, obtained from EDX measurements. (c) Calculated Cu occupancy, $x$, in LaCu$_{x}$Sb$_{2}$. Lattice parameter $a$ (d) and $c$ (e) of LaCu$_{x}$Sb$_{2}$ as a function of Cu occupancy $x$. (f) $c/a$ ratio as a function of $x$. Lattice parameters of earlier studies in (d) and (e) are taken from: single 1 \cite{ref1}, single 2 \cite{ref2}, single 3 \cite{ref3}, poly 1 \cite{ref3}, poly 2 \cite{ref7}, poly 3 \cite{ref4}, poly 4 \cite{ref5}, poly 5 \cite{ref6}, and poly 6 \cite{ref8}.}
\label{Fig1}
\end{figure*}

We have grown single crystals of LaCu$_{x}$Sb$_{2}$ ($0.92 \leq x \le 1.12$) by controlling the initial loading compositions. The high purity La, Cu, and Sb were loaded into alumina crucibles with different molar ratios to control Cu-site occupancies, as described in Tab.~\ref{Table1}. In the table, samples \# $2-7$ were grown by different Cu/Sb ratios with fixed La concentration, while samples \# 4-1 and 4-2 were prepared by varying La concentration with fixed Cu/Sb ratio, as depicted in the ternary phase diagram of Fig.~\ref{Fig1}(a). The molar ratio for sample \# 1 follows  Ref.~\cite{ref1}. The prepared ampoule was heated to 1000 $^{o}$C and cooled down to 700 $^{o}$C for over 100 hours. After decanting excessive Cu/Sb liquid by a centrifuge, rectangular plate-like crystals were obtained as shown in Fig.~\ref{Fig1}(a). When the cooling rate is decreased to 1~$^{o}$C per hour, in some specific conditions (not always), samples can be grown to as large as $\sim 1$~gram size, as shown in Fig.~\ref{Fig1}(a).

Powder x-ray diffraction (XRD) measurements were performed at room temperature with Cu $K_{\alpha}$ radiation using a Rigaku Miniflex diffractometer. Analysis of the powder XRD patterns confirms that samples crystallize in the tetragonal HfCuSi$_{2}$-type structure ($P4/nmm$, No.129). Silicon powders as an internal standard were used to correct the instrument zero shift. The chemical compositions of the crystals were determined using a Quattro Environmental scanning electron microscope (ESEM) produced by Thermofisher.

%Figure1

Four-probe resistivity measurements were performed down to 0.4~K in a Quantum Design (QD) Physical Property Measurement System (PPMS) with He$^{3}$ option, equipped with Lake Shore model 372 resistance bridge and Keithley Delta-mode resistance measurement setup. Four Pt-wires were attached on a sample using silver paste. The electrical current was applied in the $ab$-plane of the crystal ($I \parallel ab$) and the magnetic field was applied to the $c$-axis ($H \parallel c$). Specific heat was measured by the relaxation method in a QD PPMS. The dc magnetization was measured up to 70~kOe in a QD Magnetic Property Measurement System (MPMS3) with a SQUID VSM option.

The electronic structure calculations were performed using first-principles methods within the {QUANTUM} {ESPRESSO} suite~\cite{Giannozzi2009}, implementing the density functional theory (DFT). For the exchange-correlation contribution to the total energy, we used the Generalized Gradient Approximation (GGA) functional in the parameterization of Perdew-Burke-Ernzerhof (PBE)~\cite{JPPerdew1996_PRL} within a fully-relativistic scheme. The lattice parameters were fixed to the experimental values, and the atomic positions were then allowed to fully relax until the forces on the atoms became less than $10^{-4}$ Ry/a.u. Core electrons were treated with the optimized norm-conserving Vanderbilt pseudopotentials~\cite{Hamann_PRB2013}, while the valence electrons were described with plane waves up to a kinetic energy and charge density cutoff of $70$~Ry and $280$~Ry, respectively. For the self-consistent calculations, the Brillouin zone of the primitive cell was sampled with a $10\times10\times10$ Monkhorst-Pack~\cite{HJMonkhorst1976} \textbf{k}-point grid using a Marzari-Vanderbilt~\cite{Marzari99} smearing of 0.01 Ry. Finally, non-self consistent calculations were performed to obtain the density of states (DOS) and Fermi surfaces (FS) using a dense 2541 \textbf{k}-points. The FS visualization was done by utilizing the FERMISURFER package~\cite{Kawamura2019_FS}.

\section{Result and discussion}

Figure 1(b) shows the results of Energy-Dispersive x-ray (EDX) measurements, where elemental compositions of each sample were determined by both area scans and spot scans. The spot scans provide similar elemental concentrations through scanned spots, consistent with that obtained from area scans. It has to be noted that a significant local variation of atomic ratios is detected from the spot scans only for selected samples in \# 1, implying a chemical inhomogeneity and probably reaching a solubility limit. The obtained atomic ratios (\%) from the area scans are summarized in Tab.~\ref{Table1}. The area scan clearly reveals that atomic concentrations of Cu and Sb in the sample strongly depend on initial loading Cu/Sb ratios. The Cu-site occupancy, $x$, is obtained by the (3$\times$Cu)/(La+Sb) formula and plotted in Fig.~\ref{Fig1}(c) as a function of the loading Cu/Sb ratio. It can be clearly seen that the $x$ value increases from $\sim$0.92 to $\sim$1.12 as increasing Cu concentration from 3.1 \% to 47.5 \% in the flux. On the other hand, the variation of La concentration in the flux has a subtle effect on the final stoichiometry as summarized in Tab.~\ref{Table1}. The obtained $x$ values of sample \# 4, 4-1, and 4-2 are within the error range of the EDX measurements. Therefore, we infer the actual $x$ value for sample \# 4-1, and 4-2 from the residual resistivity ratio (see below the electrical resistivity result). The results of EDX analysis suggest that the Cu-site occupancy in LaCu$_{x}$Sb$_{2}$ can be controlled by varying the initial loading Cu/Sb compositions, while the variation of La composition with fixed Cu/Sb ratio does not significantly affect the Cu occupancy.

Lattice parameters $a$ and $c$ obtained from the refinement of powder XRD patterns are plotted in Fig.~\ref{Fig1}(d) and (e), respectively, as a function of the Cu-occupancy $x$ and summarized in Tab.~\ref{Table1}. Note that in addition to sample peaks the XRD pattern reveals pure Sb peaks only for sample \#1 ($x$ = 0.92), while no Sb or secondary phases are detected from samples for $x > 0.92$. This is somewhat consistent with the result of EDX that only sample \#1 shows inhomogeneous elemental concentrations. The lattice parameter $a$ is smallest for the $x$ value closest to 1 and $c$ increases up to $x = 1$ and starts to saturate toward $\sim 10.4$~\AA. The lattice parameters obtained in this study are basically consistent with earlier studies~\cite{ref1, ref2, ref3, ref4, ref5, ref6, ref7, ref8}. To eliminate instrumental errors the $c/a$ ratio in this study is compared to previous reports. It is evident from Fig.~\ref{Fig1}(f) that the $c/a$ ratio increases quasi-linearly up to  $x = 1$ and is almost independent of $x$ for $x > 1$. Note that the $c/a$ ratio of single 1~\cite{ref1} and single 2~\cite{ref2} deviates from a general trend, implying that the actual $x$ value is smaller than 1.

%Figure2 

\begin {figure}
\includegraphics[width=1\linewidth]{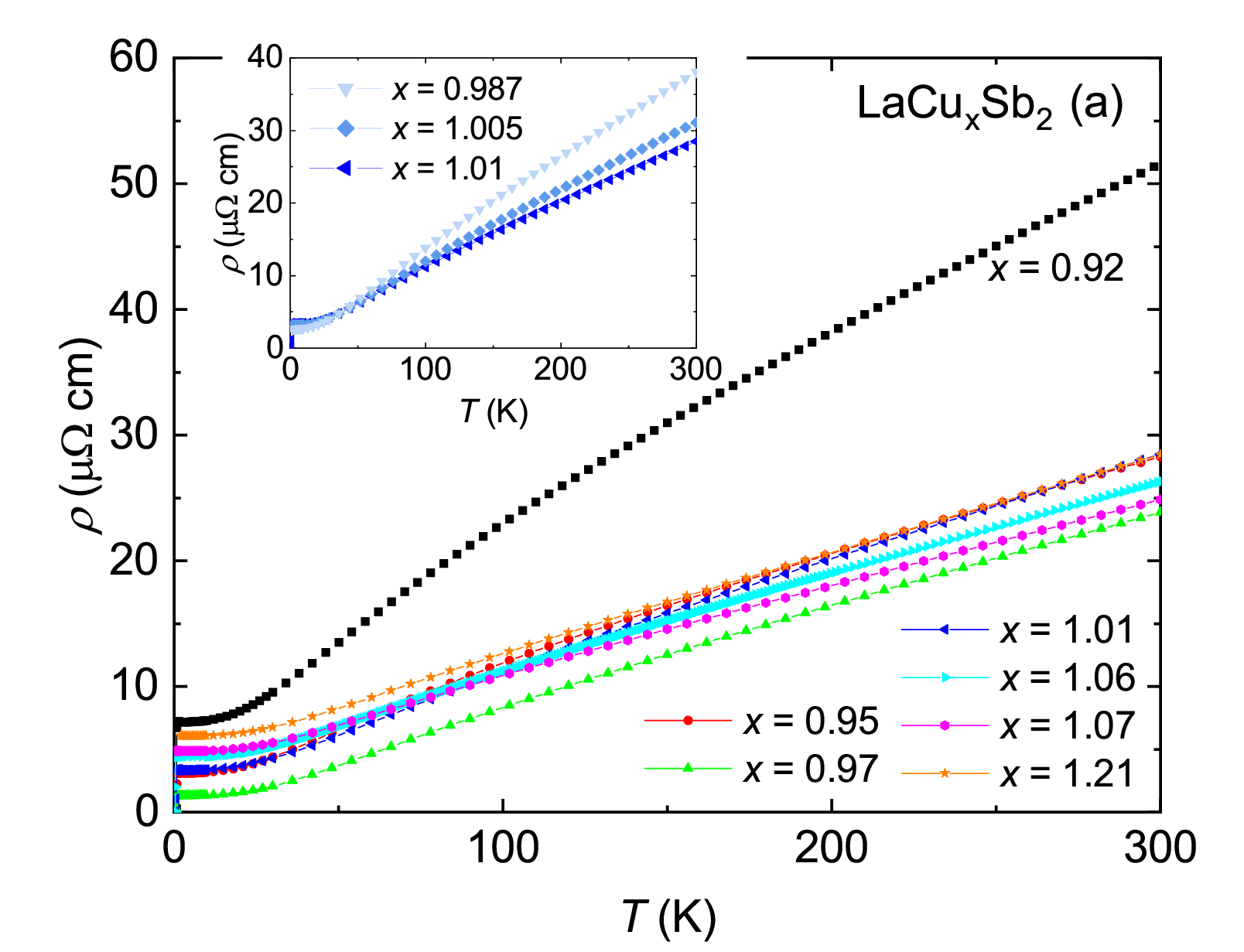}
\includegraphics[width=1\linewidth]{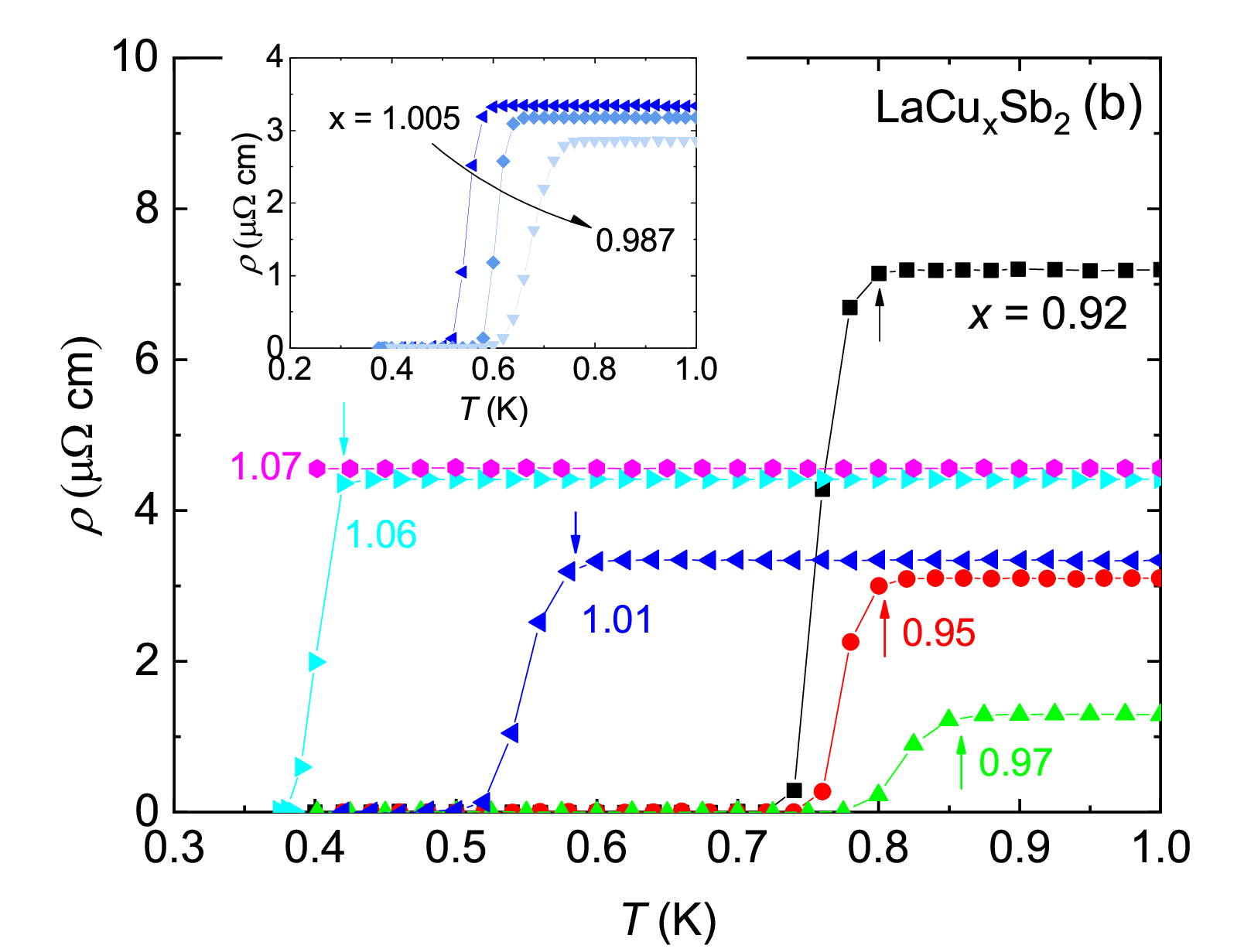}
\caption {(a) $\rho(T)$ of LaCu$_{x}$Sb$_{2}$. Inset shows $\rho(T)$ curves for $x$ = 0.987, 1.005, and 1.01. (b) $\rho(T)$ of LaCu$_{x}$Sb$_{2}$ below 1~K. Vertical arrows indicate the onset of superconducting transition temperatures. Inset shows $\rho(T)$ curves for $x$ = 0.987, 1.005, and 1.01.}
\label{Fig2}
\end{figure}

%Figure3 

\begin {figure}
\includegraphics[width=1\linewidth]{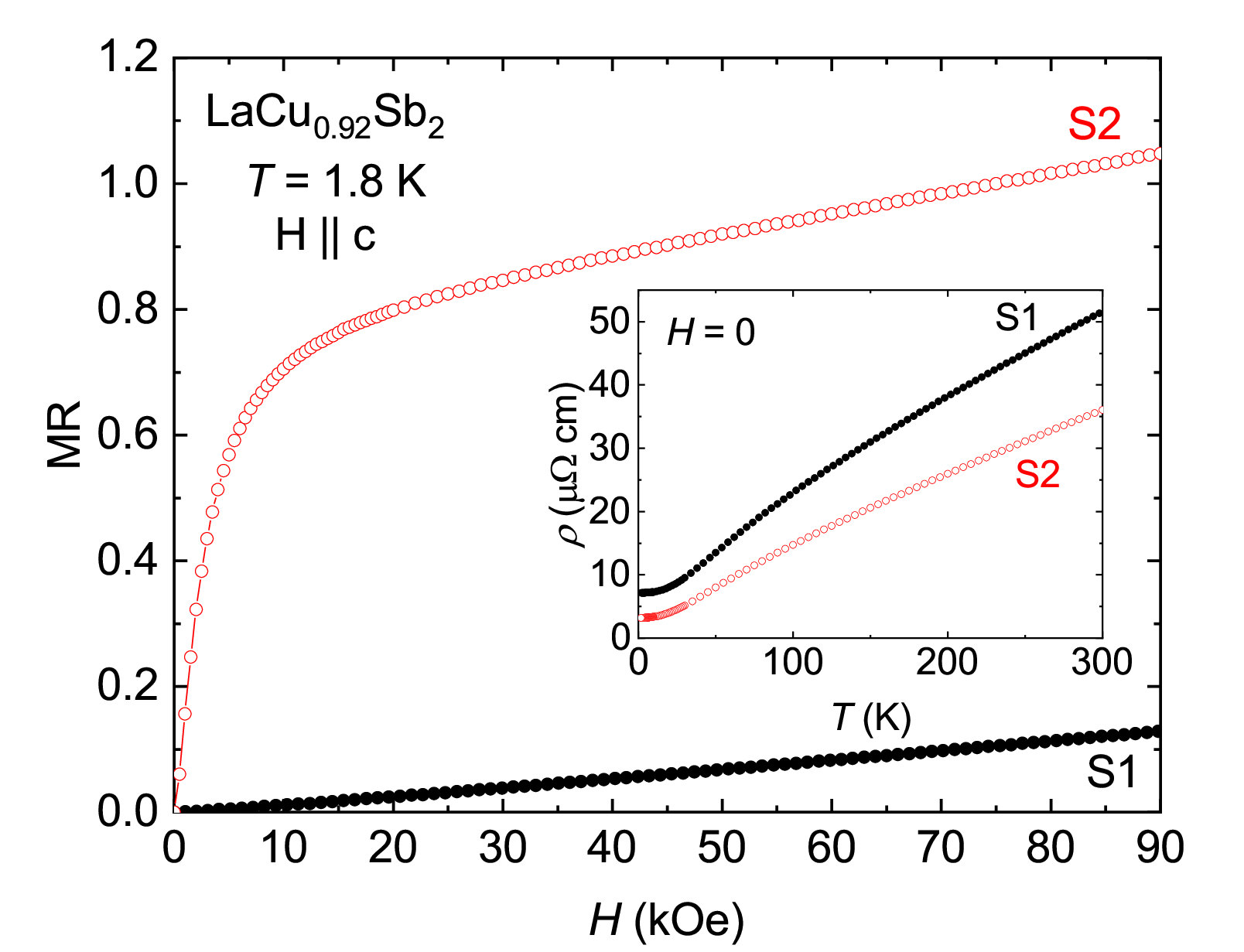}
\caption {Magnetoresistance (MR), $\rho(H)-\rho(0)/\rho(0)$, of LaCu$_{0.92}$Sb$_{2}$ at 1.8~K for two selected samples (S1 and S2) in the same batch. Inset shows $\rho(T)$ curves of S1 and S2.}
\label{Fig3}
\end{figure}

The temperature dependence of the electrical resistivity, $\rho(T)$, of LaCu$_{x}$Sb$_{2}$ is shown in Fig.~\ref{Fig2}(a). The $\rho (T)$ curves show a typical metallic behavior down to 1~K with a slight slope change below $\sim$100~K. The absolute value of the resistivity for $x$ = 0.92 is much bigger than that of other samples, which is beyond the error range of sample dimensions. Thus, we measured electrical resistivity of several pieces from the same batch and observed strong sample dependences. Figure~\ref{Fig3} shows $\rho(T)$ and magnetoresistance (MR) of two representative samples (S1 and S2). Although the temperature dependence is similar, the absolute value of the resistivity at 300~K varies within the range of $30 \sim 60 \mu\Omega$~cm. In particular, MR indicates totally different field dependences: MR of S1 is very small and follows a quasi-quadratic field dependence, whereas MR of S2 shows a large, linear field dependence at low fields followed by a linear increase at high fields. It has to be noted that both $\rho(T)$ and MR of S2 is consistent with an earlier study~\cite{ref1}. These two resistivity samples, having polished their surfaces, were further characterized by EDX. The spot scans indicate that the Cu concentration in S1 is almost constant, varying $x = 0.92~ \pm~ 0.018$. On the other hand, the spot scans of S2 clearly confirm a sample inhomogeneity with a large variation of Cu concentration (0.84 $< x <$ 0.93). Note that for a particular spot with the area ($\sim114$~$\mu$m$^{2}$) Sb concentration in S2 is detected to be 98.7 \%, implying an inclusion of fluxes or substitutional solid solutions. This EDX result for $x$ = 0.92 suggests that the Sb peaks in XRD patterns are mostly due to the flux inclusions. It is evident from powder XRD, EDX, and resistivity measurements that the large linear MR is not intrinsic property of the sample but extrinsic effect.

%Figure4 

\begin {figure}
\includegraphics[width=1\linewidth]{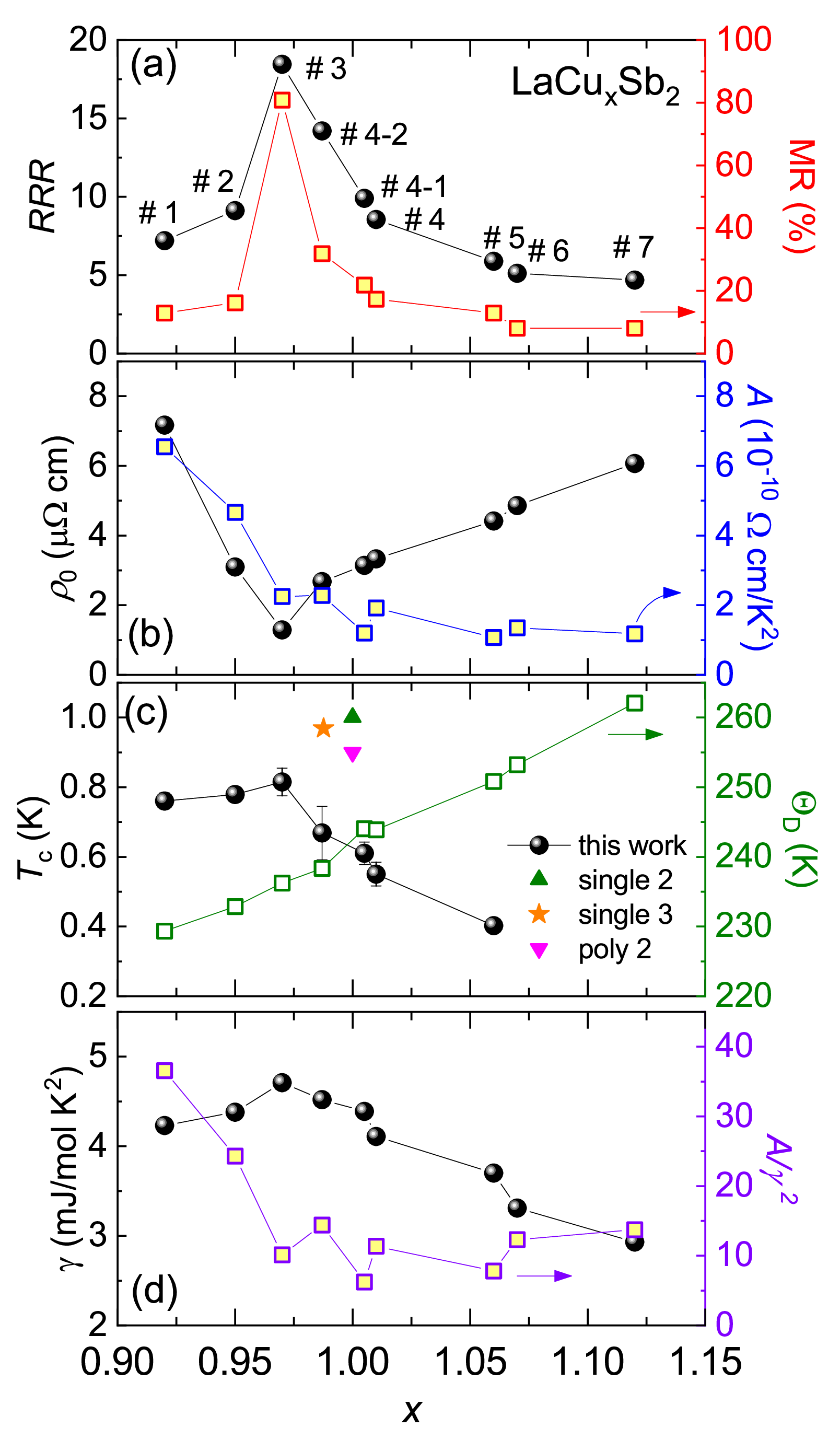}
\caption{Parameters obtained from resistivity $\rho(T)$ and specific heat $C_{p}(T)$ measurements, plotted as a function of Cu occupancy $x$. (a) Residual resistivity ratio (RRR) and magnetoresistance (MR) at 1.8 K and 90 kOe. (b) Residual resistivity $\rho_{0}$ and the coefficient $A$ of $T^2$ dependence. (c) Superconducting transition temperature $T_{c}$ determined from $\rho(T)$ and Debye temperature $\Theta_{D}$ obtained from $C_{p}(T)$. $T_{c}$ values of earlier studies are taken from: single 2 \cite{ref2}, single 3 \cite{ref3}, and poly 2 \cite{ref4}. (d) Electronic specific heat coefficient $\gamma$ and Kadowaki-Wood ratio in the unit of $\mu\Omega$ cm mol$^{2}$ K$^{2}$ J$^{-2}$.}
\label{Fig4}
\end{figure}

The residual resistivity ratio (RRR), determined by $\rho$(300 K)/$\rho$(1.8 K), is plotted in Fig.~\ref{Fig4}(a). RRR shows a clear $x$-dependence, with the highest value of RRR $\sim$ 16 at $x$ = 0.97. Since EDX measurements do not provide enough resolutions to distinguish Cu concentrations of sample \#4-1 and 4-2, the $x$ values of these samples are inferred from a linear interpolation of the RRR values between sample \#3 and \#4 to be $x$ = 1.005 for sample \#4-1 and $x$ = 0.987 for sample \#4-2. $\rho(T)$ curves for $x$ = 0.987 and 1.005, together with $x = 1.01$, are plotted in the inset of Fig.~\ref{Fig2}(a). At low temperatures, the $\rho (T)$ curves of LaCu$_{x}$Sb$_{2}$ are analyzed as a characteristic of a Fermi liquid, $\rho(T) = \rho_{0} + AT^{2}$. The residual resistivity, $\rho_{0}$, and coefficient of $T^{2}$ dependence are determined by fitting the curve below 10~K. $\rho_{0}$ shows a strong $x$-dependence with the lowest value of $\rho_{0}$ = 1.29 $\mu\Omega$ cm at $x$ = 0.97, as illustrated in Fig.~\ref{Fig4} (b). The obtained $A$ value drops sharply as $x$ increases to 0.97, while for $x \ge 1$ $A$ weakly depends on $x$.

Below 1~K $\rho (T)$ curves indicate a superconductivity for $x \le 1.06$, plotted in Fig.~\ref{Fig2}(b), consistent with earlier reports \cite{ref2, ref3, ref7}. The superconducting transition temperature, $T_{c}$, is determined as the temperature at middle resistivity value between normal state (onset criteria) and zero resistance (offset criteria). The determined $T_{c}$ is plotted in Fig.~\ref{Fig4}(c) as a function of $x$, including $T_{c}$ values taken from Refs.~\cite{ref2, ref3, ref7}. The upper and lower limit of error bars represent $T_{c}^{onset}$ and $T_{c}^{zero}$, respectively. $T_{c}$ increases from 0.76 K for $x$ = 0.92 to 0.82~K for $x$ = 0.97, and then decreases to 0.38~K for $x$ = 1.06. No superconductivity is observed down to 0.36~K for $x \ge 1.07$. Although the superconductivity is observed in a wide range of $x$, the highest $T_{c}$ value in this study is somewhat lower than that of earlier studies. Since physical properties of LaCu$_{x}$Sb$_{2}$ are very sensitive to Cu-site occupancies, slightly different $x$ and sample quality might lead to inconsistent $T_{c}$ values. Our results and previous studies illustrate that $T_{c}$ of LaCu$_{x}$Sb$_{2}$ strongly depends on the Cu stoichiometry.

%Figure5 %%%%%%%%%%%%%%%%%%%%%%%%%%%%%%%%%%%%%%%%%%%%%%%%%%%%%%%%%%%%%%%%%
\begin {figure}
\includegraphics[width=1\linewidth]{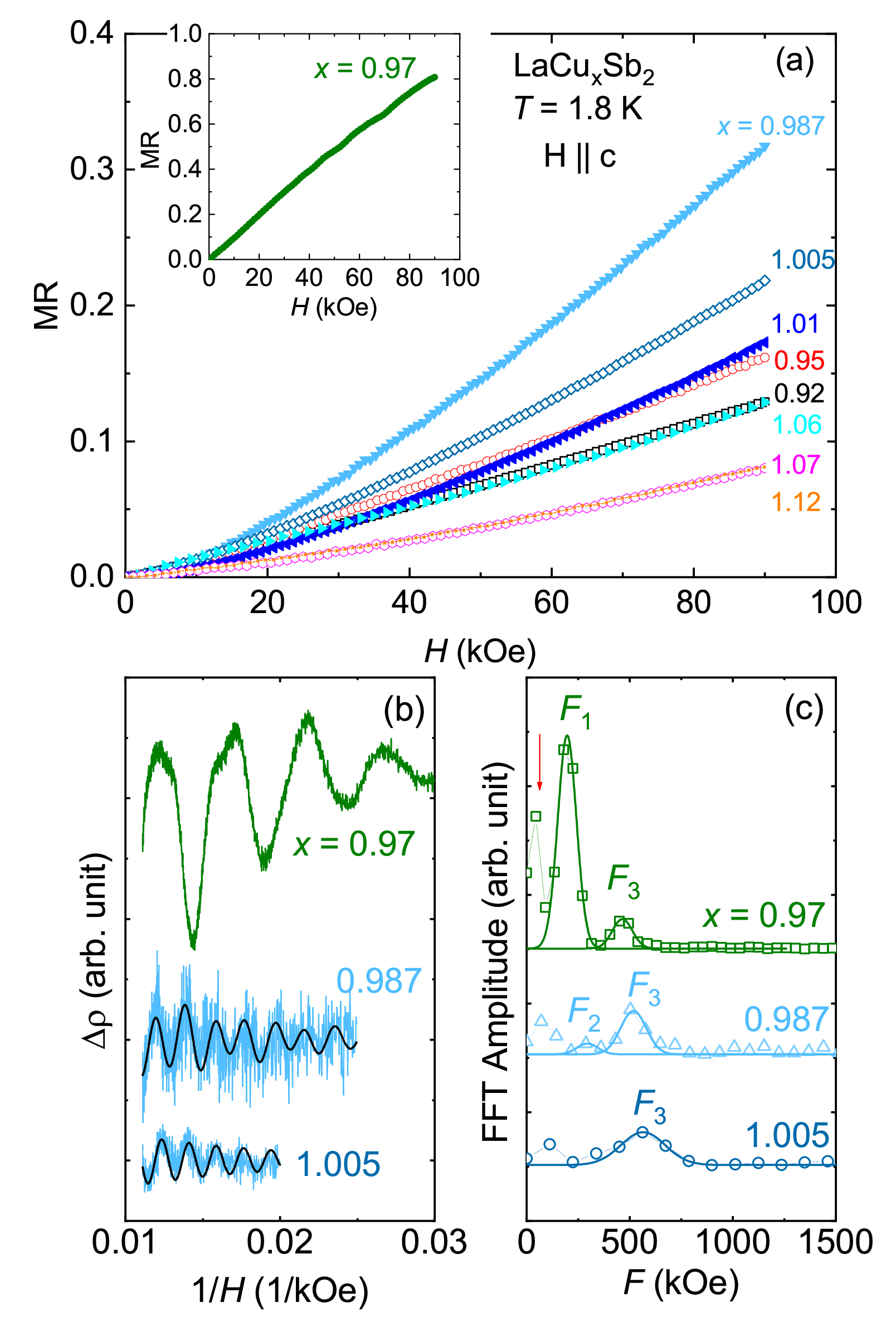}
\caption{Magnetoresistance (MR) of LaCu$_{x}$Sb$_{2}$ at 1.8~K for $H \parallel c$. Inset shows MR at $x = 0.97$. (b) SdH oscillations for $x$ = 0.97, 0.987, and 1.005, plotted after subtracting a background. Solid lines are calculated curves by using frequencies obtained from FFT. (c) Open symbols are results of FFT spectra and solid lines represent fit curves by Gauss function. The peak at low fields, indicated by vertical arrow, is an artifact of FFT.}
\label{Fig5}
\end{figure}

Figure~\ref{Fig5}(a) shows the magnetoresistance, MR = $\rho(H) - \rho(0)/ \rho(0)$, curves of LaCu$_{x}$Sb$_{2}$ at 1.8 K for $H \parallel c$. Unlike one of the earlier studies \cite{ref1} MR indicates a quadratic field dependence for most of the samples studied. For $x$ = 0.97 a remarkably large value of MR with a quasi-linear field dependence is observed as shown in the inset. The MR value at 90 kOe and 1.8 K, plotted in Fig.~\ref{Fig4}(a), also shows a strong $x$ dependence with the largest value at $x$ = 0.97, which is basically the same as RRR variation. Interestingly, although the oscillation amplitude is small, the Shubnikov-de Haas (SdH) oscillations are detected for $x$ = 0.97, 0.987, and 1.005. SdH oscillations can be clearly seen after subtracting the background MR using 2nd order polynomial fit, where the oscillations are periodic in $1/H$ as shown in Fig.~\ref{Fig5}(b). For $x$ = 0.97 and 0.987 two oscillation frequencies can be identified in the fast Fourier transform (FFT) spectra, while only one frequency is detected for $x$ = 1.005. Due to insufficient oscillations up to 90 kOe, the peak in FFT spectra is broad as plotted in Fig.~\ref{Fig5}(c). Thus, we determine the frequency by fitting the peak with the Gaussian function. The determined frequencies (peak positions) are summarized in Tab.~\ref{Table2}. The solid lines in Fig.~\ref{Fig5}(b) represent the calculated oscillations with frequencies of $F_{2}$ = 295~kOe and $F_{3}$ = 518~kOe for $x$ = 0.987 and $F_{3}$ = 567~kOe for $x$ = 1.005. It is worth emphasizing that the highest $T_{c}$, the largest RRR and MR, and the smallest $\rho_{0}$ value imply the sample with $x$ = 0.97 is the highest quality single crystal among all studied samples. Although it has to be confirmed by high resolution EDX measurements, it is not unreasonable to assume that the Cu-site occupancy of sample \#3 ($x$ = 0.97) is closest to the ideal stoichiometry 1.

%Table II %%%%%%%%%%%%%%%%%%%%%%%%%%%%%%%%%%%%%%%%%%%%%%%%%%%%%%%%%%%%%%%%%
\begin{table*}
\caption{Summary of quantum oscillation frequencies}

\begin{tabular}{l l c c c c c c}
\hline
\hline
~~~$x$~~~ & ~ & \multicolumn{4}{c}{$H \parallel c$}  & \multicolumn{2}{c}{$H \parallel ab$} \\

 &  &~~~$F_{1}$ (kOe)~~~ & ~~~$F_{2}$ (kOe)~~~  & ~~~$F_{3}$ (kOe)~~~ & ~~~$F_{4}$ (kOe)~~~ & ~~~$F_{1}$ (kOe)~~~ & ~~~$F_{2}$ (kOe)~~~ \\ \hline

0.92     & dHvA                       &    &     & 538    &  & 639	&        \\ \hline
0.95     & dHvA                       &    &     & 522    &  & 662	& 891 \\ \hline \\

0.97     & dHvA                       & 193                       &  393                   & 467                           & 12754                 & 160   & 619 \\
            & 2nd harmonics        &                             &  775                  & 936                           &                              &             &  \\
            & SdH                         & 197                     &                            & 465                          &                               &             &  \\        
            & DFT calculations     & $\alpha$ = 216    & $\beta$ = 554     & $\gamma$ = 658     & $\delta$ = 13194    &           &  \\ \\  \hline

0.987   & dHvA                       &                             &  340                    & 513                           &                                 & 110  & 605 \\
            & SdH                         &                             &  295                    & 518                          &                               &             &  \\ \hline

1.005   & dHvA                       &    &  301   & 572    &  &        & 785 \\
            & SdH                         &    &           & 567    &  &        &  \\ \hline
1.01     & dHvA                       &    &  287   & 617    &  &         &  \\ \hline
1.06     & dHvA                       &    &  192   & 720    &  & 527  &  \\
            & 2nd Harmonics       &    &  387   &            &  &         &  \\  \hline
1.07     & dHvA                       &    &  102   & 843    &  & 273  &  \\  \hline
1.12     & dHvA                       &    &  52     & 957    &  & 151  &  \\

\hline
\hline
 
\end{tabular}
\label{Table2}
\end{table*}

%Figure6 %%%%%%%%%%%%%%%%%%%%%%%%%%%%%%%%%%%%%%%%%%%%%%%%%%%%%%%%%%%%%%%%%
\begin {figure}
\includegraphics[width=1\linewidth]{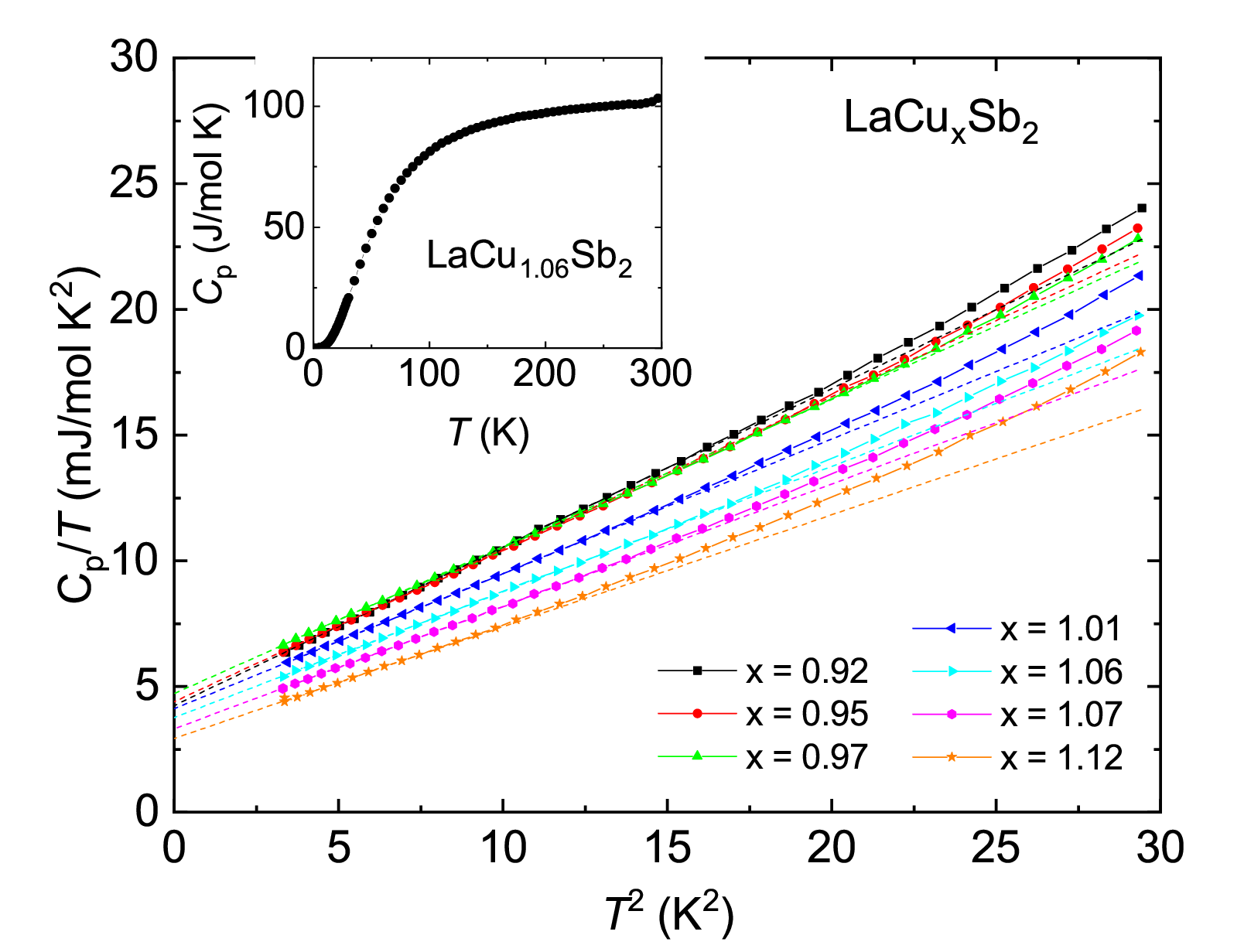}
\caption {$C_{p}/T$ vs. $T^{2}$ of LaCu$_{x}$Sb$_{2}$ below 5.5 K. Dashed lines are a guide to the eye. Inset shows $C_{p}(T)$ of LaCu$_{1.06}$Sb$_{2}$ below 300~K.}
\label{Fig6}
\end{figure}

The temperature dependence of the specific heat, $C_{p}(T)$, of LaCu$_{x}$Sb$_{2}$ follows a typical behavior of paramagnetic or diamagnetic intermetallic compounds. The $C_{p}(T)$ curve for $x$ = 1.06 is plotted in the inset of Fig.~\ref{Fig6} as a representative data. The absolute value of $C_{p}$ at 300~K is close to the Dulong-Petit limit and no signature of phase transition is observed down to 1.8~K for all $x$. At low temperatures, $C_{p}$ curves for all $x$ are well described by electronic and phonon contributions, $C_{p} = \gamma T + \beta T^{3}$, as shown in Fig.~\ref{Fig6}. The electronic $\gamma$ and phonon $\beta$ contribution can be estimated from $C_{p}/T$ vs $T^{2}$ plot and the obtained Debye temperature $\Theta_{D}$ and $\gamma$ are summarized in Fig.~\ref{Fig4}(c) and (d), respectively. It has to be emphasized that $\Theta_{D}$ increases almost linearly from 229 to 262~K with an increase in $x$, whereas $T_{c}$ shows a maximum at $x = 0.97$, suggesting unconventional superconductivity in LaCu$_{x}$Sb$_{2}$. The obtained $\Theta_{D} = 231$~K for $x = 0.97$ is consistent with an earlier single crystal study \cite{ref3}, but it is much higher than that of polycrystalline sample ($\Theta_{D} = 151$~K) \cite{ref7}. The obtained $\gamma$ value indicates a maximum at $x = 0.97$ and the value of $\gamma$ = 4.71 mJ/mol K$^{2}$ is consistent with earlier studies \cite{ref2, ref3}. The Kadowaki-Woods ratio ($A/\gamma^{2}$), plotted in Fig.~\ref{Fig4}(d), is close to 10 $\mu\Omega$ cm K$^{2}$ mol$^{2}$ mJ$^{-2}$ for $x \ge 0.97$, while it becomes larger for $x < 0.97$ mainly caused by the unusually large value of $A$. It is unclear the origin of the large $A$ value for $x < 0.97$. If the large $A$ value is simply due to the disorder effect, a similar enhancement can be expected for samples with $x > 1$.

%Figure7 %%%%%%%%%%%%%%%%%%%%%%%%%%%%%%%%%%%%%%%%%%%%%%%%%%%%%%%%%%%%%%%%%
\begin {figure}
\includegraphics[width=1\linewidth]{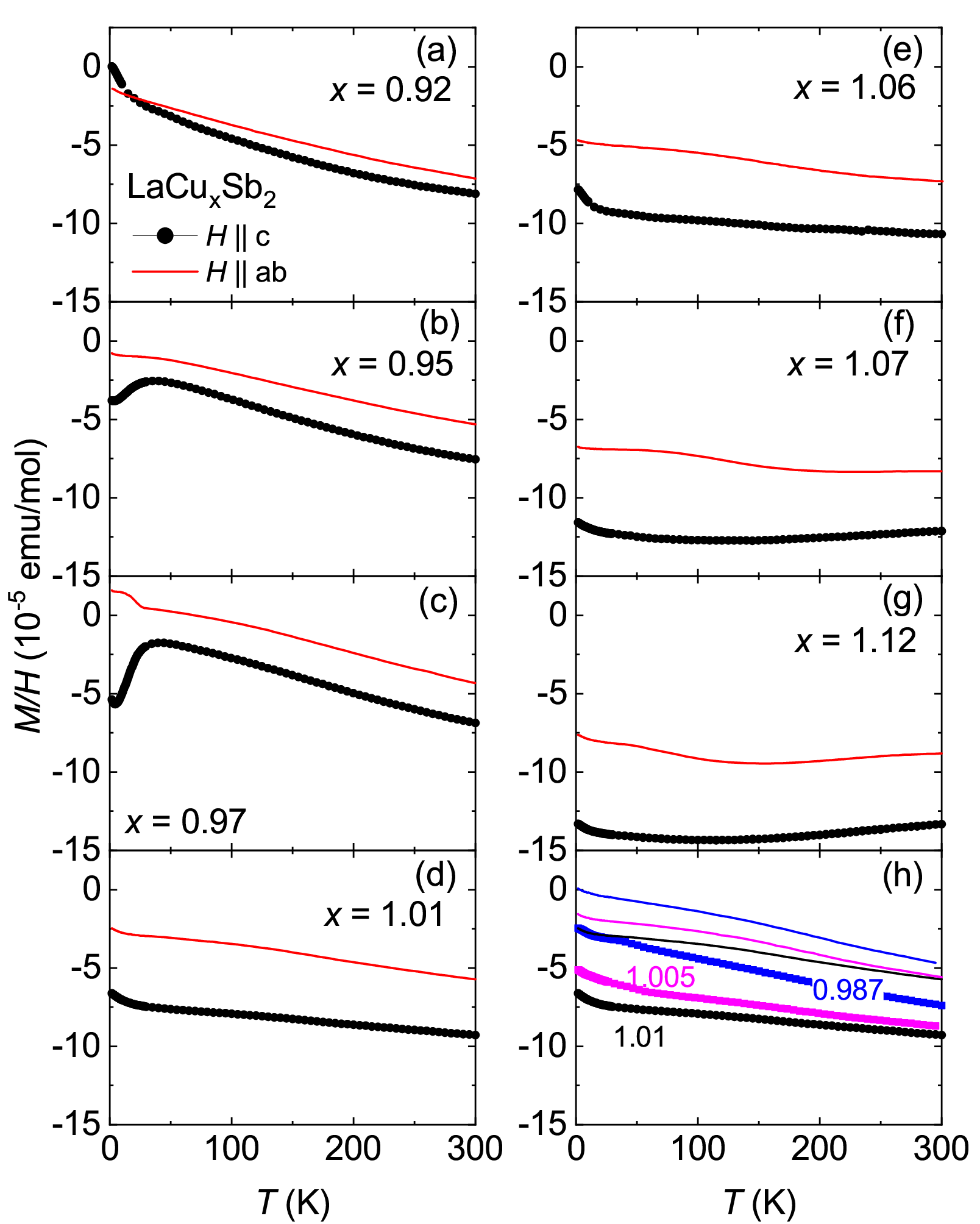}
\caption{(a-h) $M$/$H$ of LaCu$_{x}$Sb$_{2}$ at $H$ = 70 kOe for both $H \parallel c$ and $H \parallel ab$.}
\label{Fig7}
\end{figure}

%Figure8 %%%%%%%%%%%%%%%%%%%%%%%%%%%%%%%%%%%%%%%%%%%%%%%%%%%%%%%%%%%%%%%%%
\begin {figure}
\includegraphics[width=1 \linewidth]{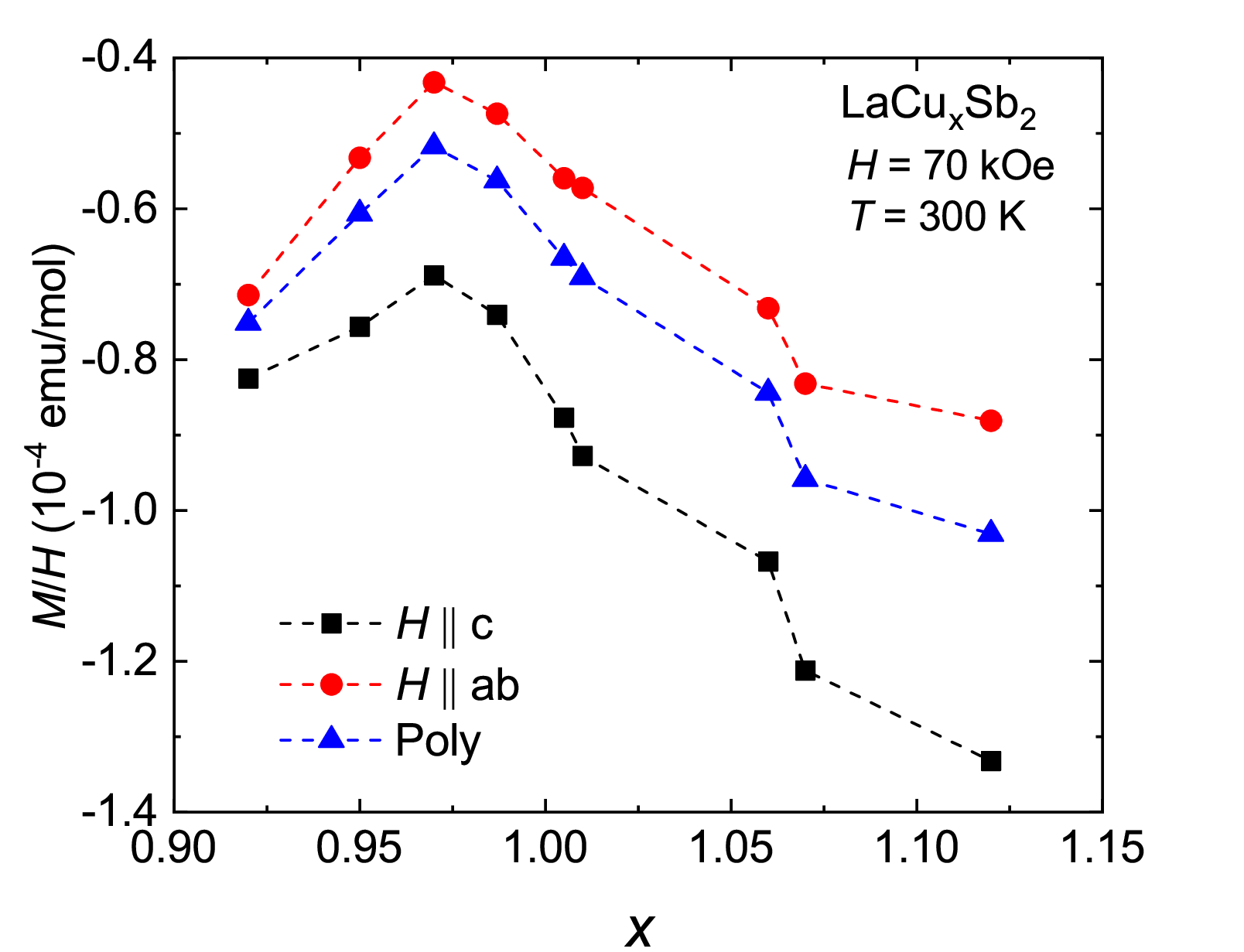}
\caption{$M/H$ values at 70~kOe and 300~K as a function of $x$. The polycrystalline averaged value is obtained by $\chi_{poly} = 2/3 \chi_{ab} + 1/3 \chi_{c}$.} 
\label{Fig8}
\end{figure}

Figures~\ref{Fig7}(a)-(h) show the temperature dependence of the magnetic susceptibility, $M/H$, curves of LaCu$_{x}$Sb$_{2}$, taken at $H = 70$~kOe for both $H \parallel ab$ and $H \parallel c$. $M/H$ curves depend weakly on temperature and indicate no signature of phase transitions down to 1.8~K. Note that at low temperatures a sudden drop or sharp rise in $M/H$ is due to the quantum oscillations (see below the magnetization isotherms). $M/H$ curves show a quasi linear temperature dependence for $x \le 1.06$, while $M/H$ curves develop a broad hump for higher $x$ values. $M/H$ curves indicate a weak anisotropy between $H \parallel ab$ and $H \parallel c$. The smallest anisotropy is observed at the smallest $x = 0.92$ and the anisotropy becomes larger as $x$ increases. For all $x$, the absolute value of the magnetic susceptibility for $H \parallel c$ is bigger than that for $H \parallel ab$, which can be clearly seen from $M/H$ value at 300~K (Fig.~\ref{Fig8}).

The observed magnetic susceptibility of LaCu$_{x}$Sb$_{2}$ deviates from a temperature independent Pauli paramagnetic or Landau diamagnetic behavior. One possible origin of this deviation can be due to contributions from paramagnetic impurities. However, the paramagnetic impurity contribution strongly affects the magnetic susceptibility curve mostly in low temperature regime. The systematic change of the magnetic susceptibility and the variation of the anisotropy as a function of $x$ rule out this possibility. In addition, $M/H$ curves for $x$ = 0.987, 1.005, and 1.01 exclude the possibility of paramagnetic impurity contribution, where $M/H$ curves continuously shift downward for entire temperatures as shown in Fig.~\ref{Fig7}(h). Note that since the sample at $x$ = 0.987 contains the highest La concentration in the growth (12\% La), the most significant impurity contribution is expected from this sample among all studied batches. Therefore, the systematic change of $M/H$ curves as a function of $x$ is expected to be due to either Landau diamagnetic contribution gets the smallest or Pauli paramagnetic contribution becomes the strongest at $x$ = 0.97. Although further detailed electronic structure calculations are necessary to explain the observed variations, the $x$-dependent $M/H$ suggests a systematic change of the electronic band structures as Cu-site occupancy varies.

%Figure9 %%%%%%%%%%%%%%%%%%%%%%%%%%%%%%%%%%%%%%%%%%%%%%%%%%%%%%%%%%%%%%%%%
\begin {figure}
\includegraphics[width=1 \linewidth]{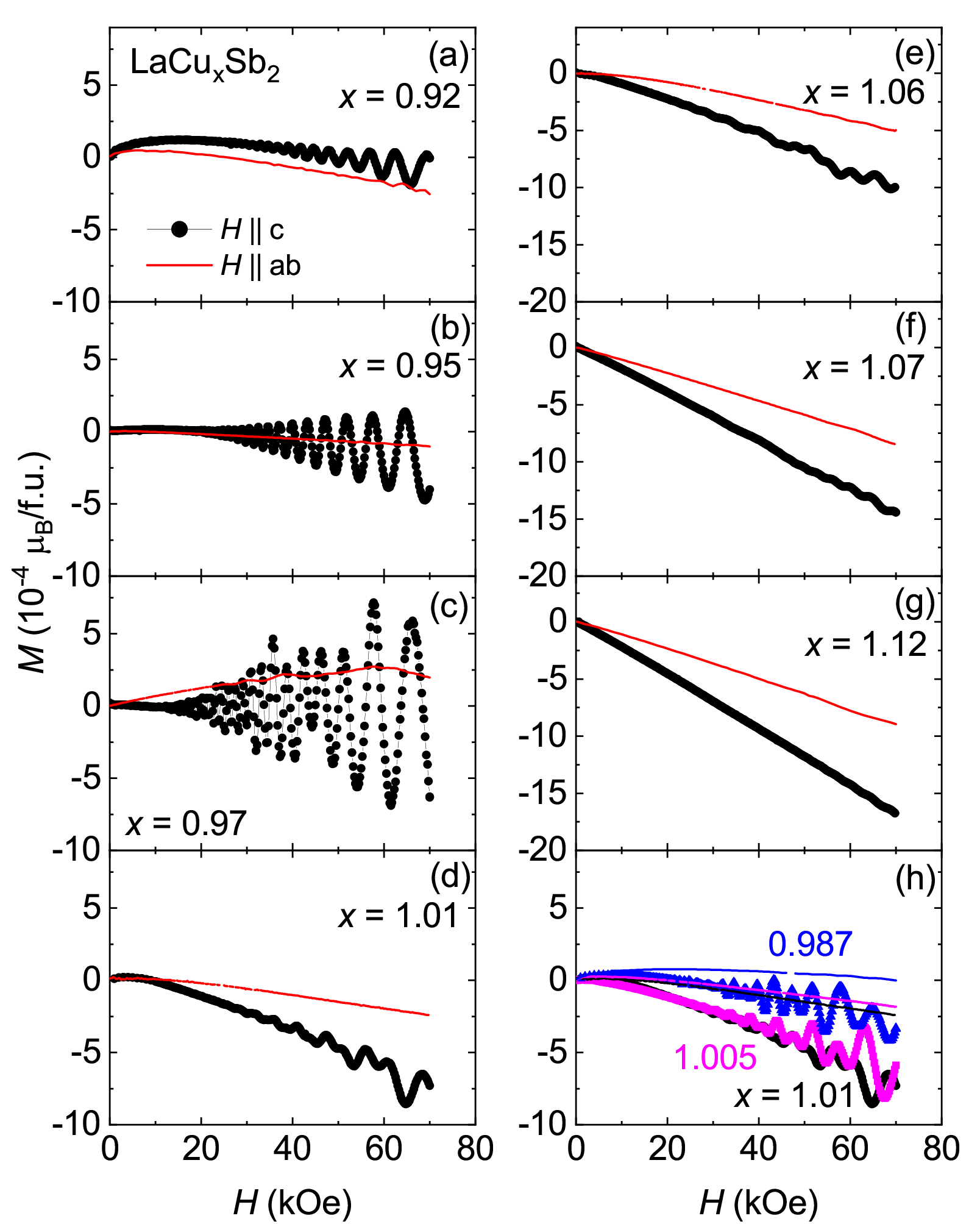}
\caption{(a-h) Magnetization isotherms of LaCu$_{x}$Sb$_{2}$ for both $H \parallel c$ (symbols) and $H \parallel ab$ (lines) at 1.8 K.}
\label{Fig9}
\end{figure}

%Figure10 %%%%%%%%%%%%%%%%%%%%%%%%%%%%%%%%%%%%%%%%%%%%%%%%%%%%%%%%%%%%%%%%%
\begin {figure*}
\includegraphics[width=1 \linewidth]{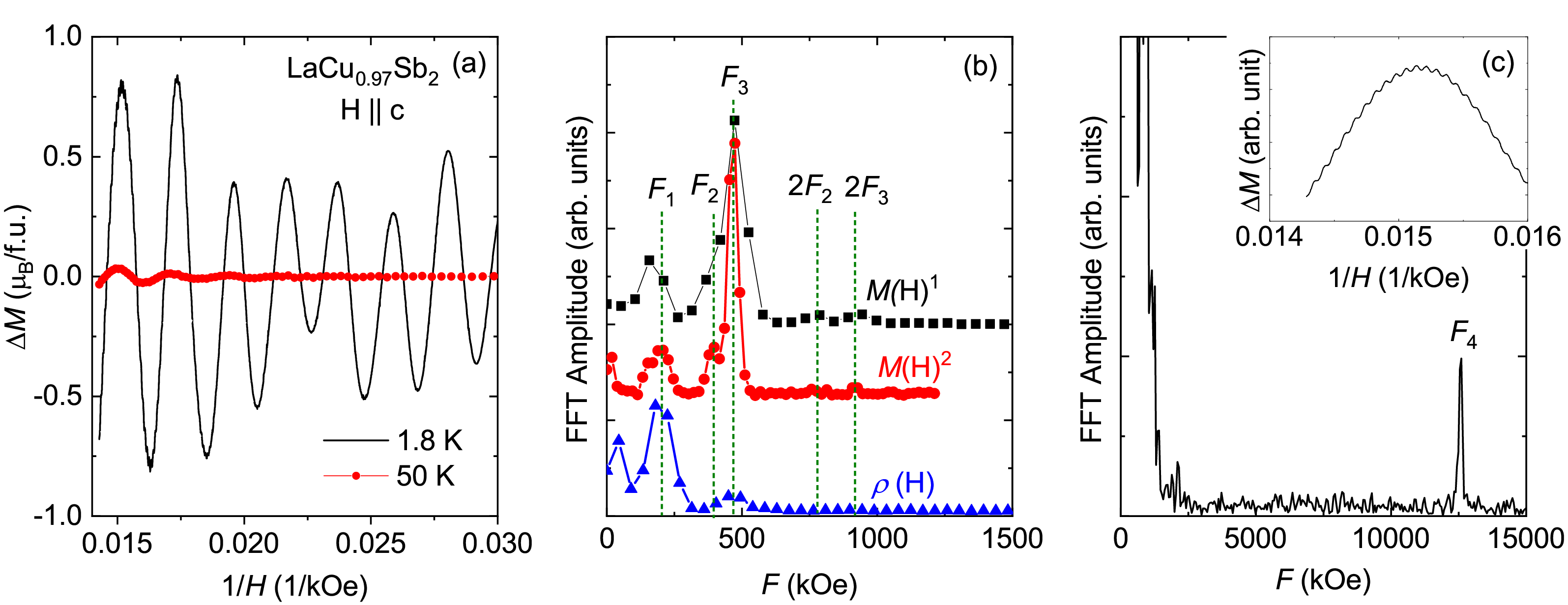}
\caption{(a) dHvA quantum oscillations in LaCu$_{0.97}$Sb$_{2}$ for $H \parallel c$ at 1.8 and 50 K. (b) FFT spectra obtained from $M(H)$ and $\rho(H)$. $M$(H)$^{1}$ and $M$(H)$^{2}$ are obtained from different field ranges (see details in text). (c) FFT spectra obtained from high field regime. Inset shows $M(H)$ above 60~kOe, plotted in $1/H$.}
\label{Fig10}
\end{figure*}

Magnetization isotherms, $M(H)$, of LaCu$_{x}$Sb$_{2}$ at $T$ = 1.8~K for both $H \parallel ab$ and $H \parallel c$ are plotted in Fig.~\ref{Fig9}. The anisotropy between $H \parallel ab$ and $H \parallel c$ is smallest at the smallest  $x$ value and becomes bigger as $x$ increases. For $H \parallel c$ prominent de Haas-van Alphen (dHvA) quantum oscillations are detected from all studied samples, where oscillations are superimposed on a mostly diamagnetic background. The oscillation amplitude is strongest for $x = 0.97$ and the oscillation amplitude decreases as $x$ decreases or increases from $x = 0.97$.

In particular for $x = 0.97$, dHvA oscillations for $H \parallel c$ are clearly observed in magnetic field as low as 10~kOe at 1.8~K and are discernible for temperatures as high as 50~K, indicating a very high quality of the sample used in the present study. The dHvA oscillations at $T$ = 1.8 and 50~K, after subtracting the background, are plotted in Fig.~\ref{Fig10}(a). It should be emphasized that the exclusively high frequency oscillation is superimposed on sinusoidal low frequency oscillations, as depicted in the inset of Fig.~\ref{Fig10}(c). Four frequencies can be clearly identified in the FFT spectra, shown in Fig.~\ref{Fig10}(b) and (c), corresponding to $F_{1} = 193$~kOe, $F_{2} = 393$~kOe, $F_{3} = 467$~kOe, and $F_{4} = 12574$~kOe. The highest frequency $F_{4}$ peak is difficult to identify in the FFT spectra due to the large difference in the amplitude. Thus, the FFT analysis is performed on the oscillatory components of magnetization in the field range of $60 \le H \le 70$~kOe, where the low frequency oscillations can be subtracted as the background. In this manner the well defined peak is clearly seen in the FFT spectra. Although the amplitude is small, the second harmonics of the frequencies $2F_{2} = 775$~kOe and $2F_{3} = 936$~kOe are also shown in the FFT spectra. The FFT spectra obtained from SdH oscillations (Fig.~\ref{Fig5}) are also plotted in Fig.~\ref{Fig10}(b), demonstrating that the frequencies are in good agreement between resistivity and magnetization measurements. We note that the oscillation frequencies are somewhat dependent on the FFT windowing and the window width. In Fig.~\ref{Fig10}(b), $M(H)^{1}$ represents the result of FFT between 30 and 70~kOe; $F_{1}$ and $F_{3}$ are clearly resolved, while $F_{2}$ is obscured by its combination of $F_{2}$ and $F_{3}$ peaks. On the other hand, $M(H)^{2}$, representing the result of FFT execution between 15 and 70~kOe, clearly resolves three peaks $F_{1}$, $F_{2}$, and $F_{3}$. The second harmonics of $2F_{2}$ and $2F_{3}$ are resolved in both $M(H)^{1}$ and $M(H)^{2}$ cases. The frequency $F_{1}$ is determined to be 168~kOe from $M(H)^{1}$ and 193~kOe from $M(H)^{2}$. The frequency obtained from $M(H)^{2}$ analysis matches well with the one obtained from the resistivity measurements ($197$~kOe). The small difference between these two measurements may be due to a slightly different sample alignment with respect to the magnetic field. The frequency $F_{3} = 467$~kOe at $x = 0.97$ in this study is consistent with the one detected from previous transport measurements \cite{ref1, ref2, ref3}, whereas other frequencies observed in this work have not been identified in earlier studies. Note that the FFT spectra in Ref.\cite{ref1, ref2, ref3} indicate a weak peak in low frequency and a small bump in high frequency regime, where we conjecture that these peaks might not be an artifact of FFTs, but correspond to actual oscillations. 

%Figure11 %%%%%%%%%%%%%%%%%%%%%%%%%%%%%%%%%%%%%%%%%%%%%%%%%%%%%%%%%%%%%%%%%
\begin {figure*}
\includegraphics[width=1\linewidth]{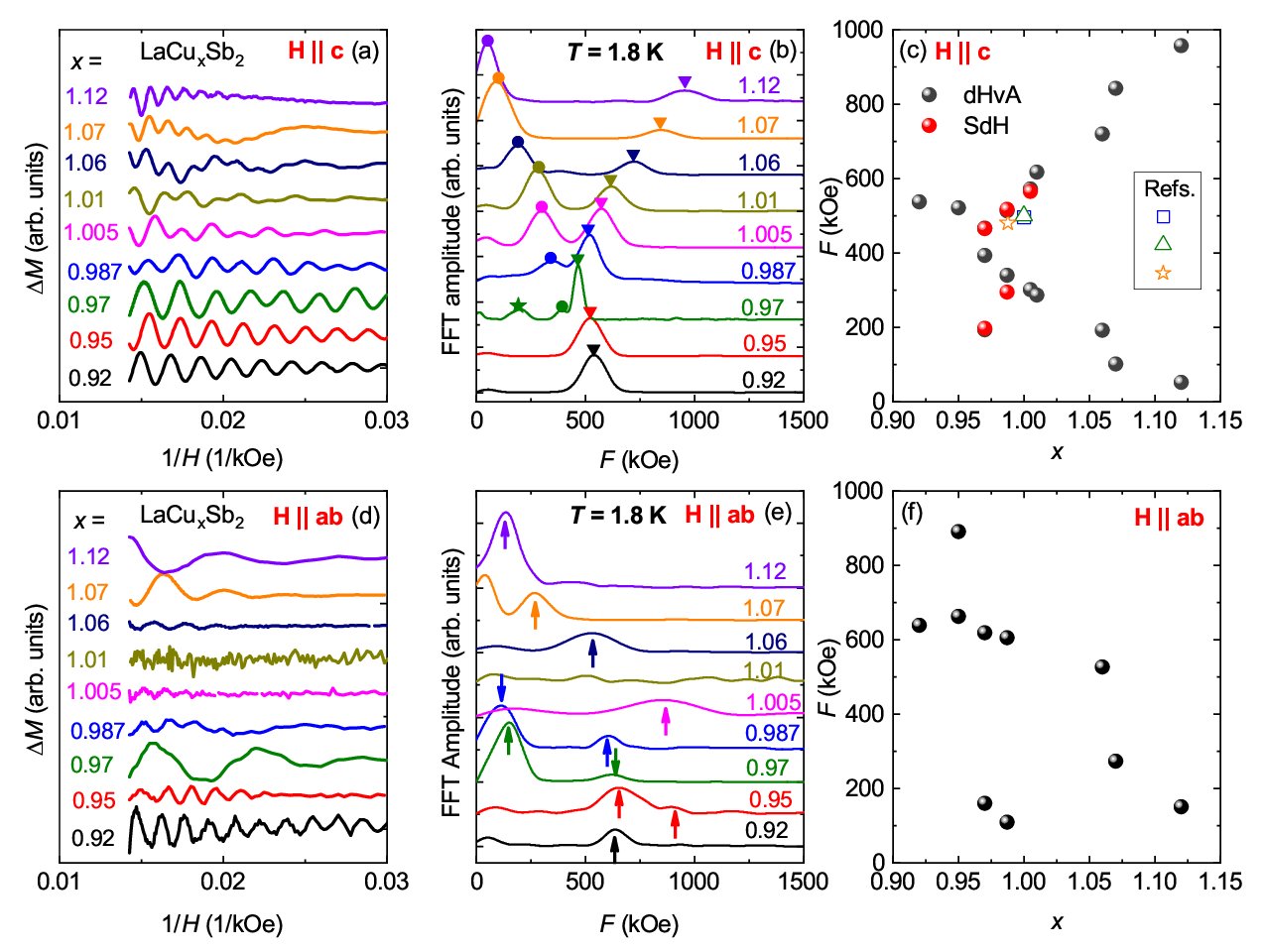}
\caption{$H \parallel c$: (a) dHvA quantum oscillations detected in LaCu$_{x}$Sb$_{2}$ at $T = 1.8$~K. (b) FFT spectra for all $x$. Symbols represent the peak positions. (c) Determined frequencies as a function $x$. Open symbols represent frequencies obtained from earlier studies \cite{ref1, ref7, AuCusub-LaAgSb2}. $H \parallel ab$: (d) dHvA quantum oscillations observed in LaCu$_{x}$Sb$_{2}$ at $T = 1.8$~K. Arrows indicates determined peak positions. (e) FFT spectra. (f) Determined frequencies.}
\label{Fig11}
\end{figure*}

For $H \parallel c$ magnetization isotherms of LaCu$_{x}$Sb$_{2}$ at $T$ = 1.8~K, after subtracting the background, are plotted in Fig.~\ref{Fig11}(a) as a function of $1/H$  and the corresponding FFT spectra are plotted in Fig.~\ref{Fig11}(b). In addition, the frequencies determined from FFT spectra, together with frequencies taken from Refs.~\cite{ref1, ref2, ref3}, are plotted in Fig.~\ref{Fig11}(c). The oscillation frequencies observed in this study are listed in Tab.~\ref{Table2}. The oscillation frequencies indicate a systematic $x$ dependence at least in a qualitative way, implying a systematic changes of the electronic structure as a function of Cu-occupancy. The main oscillation frequency near $\sim$520~kOe for $x$ = 0.92 and 0.95 seems to split into two frequencies for $x \ge 0.97$.

Although it is not pronounced in $M(H)$ for $H \parallel ab$, dHvA oscillations are discernible after subtracting the background by a polynomial fit as shown in Fig.~\ref{Fig11}(d). Note that dHvA oscillations for $H \parallel ab$ have not been reported in any of the earlier studies. The FFT results for $0.92 \le x \le 1.12$ are plotted in Fig.~\ref{Fig11}(e). The peak in low frequency ($F \sim 50$~kOe) is an artifact of the FFT analysis stemming from limited number of oscillations and very weak amplitude. Two frequencies can be identified from FFT spectra for $0.95 \le~x~\le 0.987$, while only one frequency can be obtained for $x =$ 0.92, 1.005, 1.06, 1.07, and 1.12. Note that for $x$ = 1.01 (sample \#4) we were not able to obtain the frequency due to the extremely small oscillation amplitude. In order to determine the oscillation frequency, magnetization measurements are required in either magnetic fields above 70~kOe or temperatures much lower than 1.8~K. The determined oscillation frequencies for $H \parallel ab$ are plotted in Fig.~\ref{Fig11}(f).

%Figure12 %%%%%%%%%%%%%%%%%%%%%%%%%%%%%%%%%%%%%%%%%%%%%%%%%%%%%%%%%%%%%%%%%
\begin{figure*}
\includegraphics[width=1\linewidth]{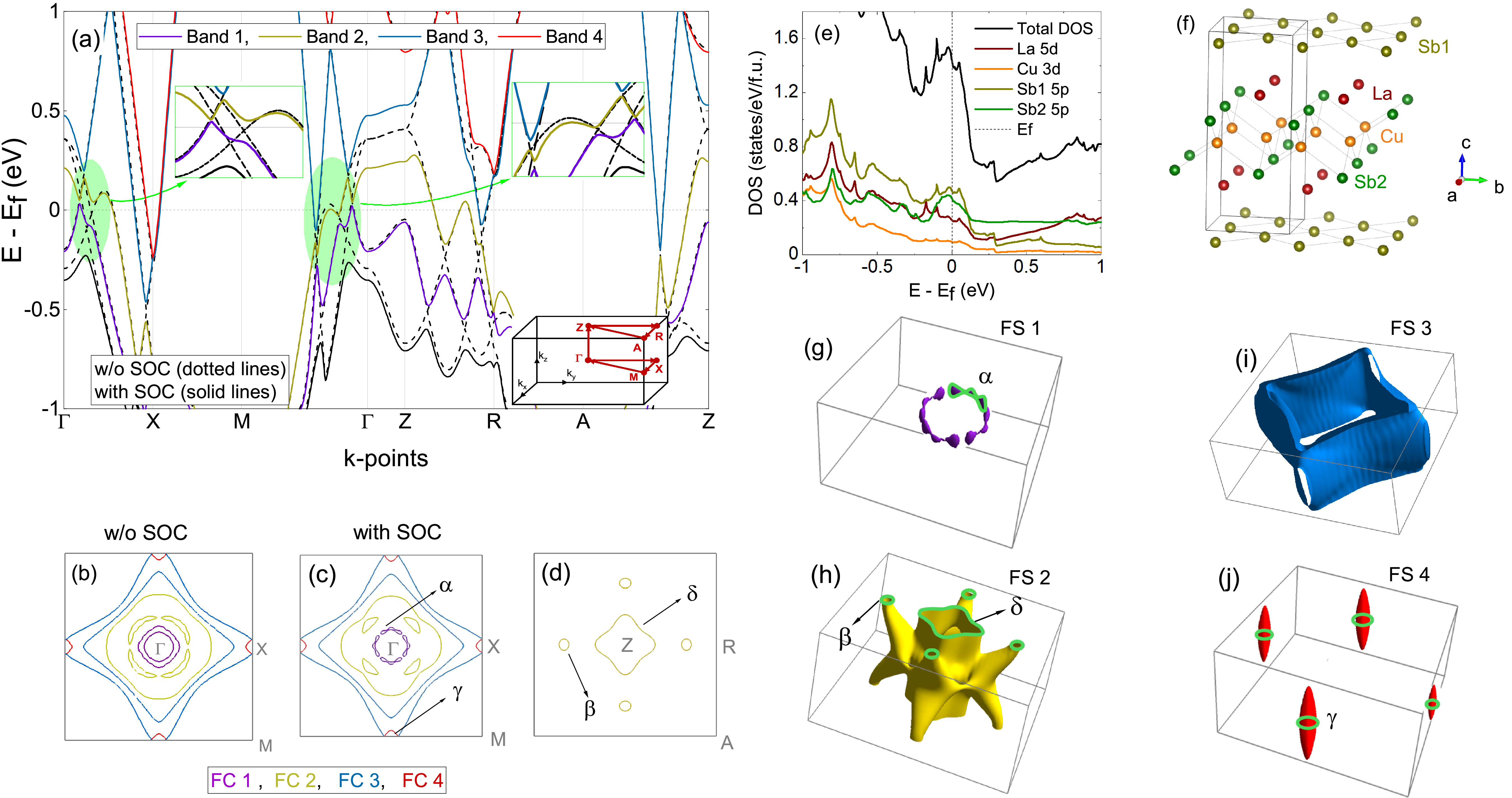}
\caption[Electronic structure of LaCuSb$_2$.]{(a) Electronic band dispersion of LaCuSb$_{2}$ for both with and without SOC. Four bands cross the Fermi level, where the marked green spots indicate clear effects of SOC near $E_f$. Fermi contours within $k_z = 0$ plane in the absent of SOC (b), and in the presence of SOC, the area of the $\alpha$ and $\gamma$ frequencies correspond to $216$~kOe and $658$~kOe respectively (c). (d) Fermi contours at the $k_z$ BZ boundary, a view along the $z-$direction. The area of the $\beta$ and $\delta$ frequencies correspond to $554$~kOe and $13194$~kOe respectively, within the SOC scheme. (e) Density of states and orbital projected density of states based on SOC calculations. (f) Crystal structure of LaCuSb$_{2}$ showing the Sb1 square net. (g) FS 1 gives rise to the $\alpha$ frequency. (h) FS 2 gives rise to both the $\beta$ and $\delta$ frequencies. (i) FS 3 has an cross-sectional area too large to be detected by our current experimental probe. (j) FS 4 gives rise to the $\gamma$ frequency. }
\label{FIG.12}
\end{figure*} 
 
Structural, thermodynamic, and transport property measurements of LaCu$_x$Sb$_{2}$ indicate a systematic variation. In particular, for $x = 0.97$ the largest RRR and MR at 90~kOe and the smallest $\rho_{0}$ value imply a minimal disorder in Cu-site, which is in good agreement with the results of quantum oscillations. Therefore, it is not unreasonable to assume that the sample with $x = 0.97$ is very close to the stoichiometric composition LaCuSb$_{2}$. The DFT calculation based on the lattice parameters for $x = 0.97$ was performed to understand the electronic structures of LaCuSb$_{2}$, where the Cu-site was assumed to be fully occupied. Figure~\ref{FIG.12}(a) shows the electronic band structure calculations with and without spin-orbit coupling (SOC) along special $k$-points in the Brillouin zone (BZ) which is shown in the inset. Calculations both with SOC and without SOC show strong linearity around point X, and along paths $\Gamma$M and AZ. This linearity has been suggested to be the source of massless electrons in LaCuSb$_{2}$~\cite{ref1, LaTESb2_Ruszala2018}. The inclusion of SOC leads to the lifting of band degeneracy far away from  $E_{f}$, notably along path $\Gamma$Z~$\sim 0.5$~eV above $E_{f}$ and path ZR~$\sim 0.5$~eV below $E_{f}$. These results are in good agreement with previous studies~\cite{LaTSb2-2014, ref1, ref2, ref3, LaTESb2_Ruszala2018}. Note that the results in Ref.~\cite{ref2} do not include SOC due to the negligible effects on bands very close to $E_{f}$.

Although there is a minimal effect of SOC on some bands within $\sim E_{f}~\pm~0.2$~eV near X and along the paths ZR, and AZ, SOC induces a pronounced lifting of the band degeneracy, indicated by the marked regions in Fig.~\ref{FIG.12}(a), leading to the band opening around $\sim E_{f}~\pm~0.2$~eV along the paths $\Gamma$X and $\Gamma$M. This lifting of the band degeneracy also induces a non-linearity of the bands along these paths rendering them parabolic, reported also in an earlier study~\cite{LaTSb2-2014}. The parabolic bands may be responsible for the near-linear behavior of the MR and Hall measurements, hence rendering this compound different from a simple Dirac system, as was indeed suggested by Ref.~\cite{ref1}. 

Figures~\ref{FIG.12}(b)~$-$~(d) show the Fermi contour (FC) plots, with the contours in Figures~\ref{FIG.12}(b) and (c) lying at the $k_z = 0$ plane calculated without and with SOC, respectively, and Figure~\ref{FIG.12}(d) is at the $k_z$ BZ boundary calculated with SOC (here without SOC plot is similar, hence only with the SOC plot is shown). Figures~\ref{FIG.12}(b) and (c) plots show four FCs, indicating that four bands cross $E_f$ in the present or absent of SOC. FC~3 and FC~4 are similar in both cases, whereas there is a drastic change in the shape and size of FC~1 when SOC is included, indicating that SOC significantly alters the Fermi surface topology. Since FC~1 with SOC gives a cyclotron frequency comparable with the one obtained from quantum oscillations, it is natural that the electronic structure (the band structure and Fermi surface) of this compound should be understood within the framework of SOC.

Figure~\ref{FIG.12}(e) shows the density of state (DOS) plot based on SOC, with the total DOS as well as the projected DOS of the atomic orbitals. The Sb1~$5p$ orbitals have major contributions to the total DOS around $E_f$, closely followed by Sb2~$5p$, and then La $5d$. Cu $3d$ has the smallest contribution. This result is in accord with previous DFT studies on this compound~\cite{ref2, LaTESb2_Ruszala2018}. In particular, this work shows that in LaCuSb$_2$, the electronic states around $E_f$ seem to be predominated by the $5p$ states from Sb1 square net, displayed in Fig.~\ref{FIG.12}(f), which is consistent with other Sb square net-based compounds such as LaAgSb$_2$ and LaAuSb$_2$~\cite{LaAgSb2-multiband, LaAuSb2-2021, LaAgSb2_Ruszala2018}. In addition, our DFT result reveals a low value for the total DOS at $E_{f}$ $:$ $1.446$~states/eV/f.u., giving $\gamma = 4.6$~mJ/mol~K$^{2}$, which is consistent with the $\gamma$ value determined from specific heat.

The Fermi surfaces (FS) based on SOC are plotted in Fig.~\ref{FIG.12}(g)~$-$~(j) corresponding, respectively, to the Bands~$1 - 4$ shown in Fig.~\ref{FIG.12}(a). Major dispersions across $E_f$ are Bands 2~$-$~4, whereas Band~1 barely crosses, indicating that the appearance of Band 1 in the FS could be sensitive to the choice of DFT parameters like the exchange-correlation functional, pseudopotential, cell dimensions and atomic positions. For example, two previous studies, based on Local Density Approximation (LDA) functional calculations do not capture the Band~1~\cite{ref1, LaTESb2_Ruszala2018}. However, a recent study using GGA functional revealed all four bands crossing $E_f$~\cite{ref2}, in agreement with the present work.

The 2D$-$contours of the FS along $k_z = 0$ plane and at the $k_z$ BZ boundary are depicted in Fig.~\ref{FIG.12}(c) and (d), respectively. In the full 3D FS structures, shown in Figs.~\ref{FIG.12}(g)~$-$~(j), FS 1 and 2 are hole bands and FS 3 and 4 are electron bands. FS 1, 3, and 4 have a quasi-2D character owing to their minimal band dispersions along $k_z$ in the BZ. FS 2 shows strong 3D character due to the presence of branch-like features roughly along $\Gamma$R path as shown in Fig.~\ref{FIG.12}(h). Such a 3D character of FS 3 has been attributed to the absence of CDW in LaCuSb$_2$ system~\cite{LaTSb2-2014}, in contrast to the CDW systems, LaAuSb$_2$ and LaAgSb$_2$ with quasi-2D FSs~\cite{LaTSb2-2014, LaAgSb2-2022, LaAgSb2_Ruszala2018}.

To understand the SdH and dHvA results for $x = 0.97$, we have calculated the frequencies associated with extremal cross-sectional areas for $H \parallel c$. Green lines in Figs.~\ref{FIG.12}(g), (h) and (j) represent cyclotron orbits, labelled as $\alpha$ (= 216~kOe), $\beta$ (= 554~kOe), $\gamma$ (= 658~kOe), and $\delta$ (= 13194~kOe). Note that FS 3 orbit is not shown, since its frequency is beyond the experimental probe in this study. In addition, FS~1 is relatively small, requiring a very dense k-point interpolation to obtain the frequency of 216~kOe, compared to the 75~kOe achieved with a sparse k-point grid.

Although DFT calculations overestimate the cyclotron orbits, the calculated frequencies are consistent with quantum oscillation measurements, suggesting that the observed SdH and dHvA frequencies $F_{1}$, $F_{2}$, $F_{3}$, and $F_{4}$ correspond to the cyclotron orbits $\alpha$, $\beta$, $\gamma$, and $\delta$, respectively. The calculated and detected frequencies are summarized in Tab.~\ref{Table2}. It must be emphasized that because FS~1 with SOC gives the closest cyclotron frequency to the observed SdH and dHvA frequency of $193$~kOe, the existence of FS~1 is significant. Furthermore, the observed major oscillatory frequency $467$~kOe arises from FS~4, in good agreement with a previous DFT study which assumed a slightly hole-doped system~\cite{ref2}. In addition to the morphology of Fermi surfaces, its cyclotron mass is another critical indicator for matching the experimentally detected quantum oscillations with electronic structure calculations. The experimentally determined and calculated effective masses of this family, together with Hall and thermoelectric power measurements, will be published elsewhere.

The Fermi surfaces of LaCuSb$_{2}$ resulting from the electronic structure calculations are similar to that reported for LaAgSb$_{2}$ and LaAuSb$_{2}$, where (except for FS 1 at BZ center) the overall shape of FS $2-4$ are very similar for all three La$T$Sb$_{2}$ ($T$ = Cu, Au, Ag) compounds. Since earlier quantum oscillation studies of LaAgSb$_{2}$ and LaAuSb$_{2}$ also detected FS orbits with very close frequencies, it is instructive to compare experimentally detected oscillation frequencies in details. The highest frequency, corresponding to Castle-starfish-shape FS 2 centered at the Z point, has been detected with oscillation frequency $\delta = 12547$~kOe for $T$ = Cu, $F = 17023$~kOe for $T$ = Au \cite{LaAuSb2_osillation}, and $F = 1290$~kOe for $T$ = Ag \cite{LaAgSb2 de Hasas-van Alphen}. The cigar-shape (or ellipsoidal-shape) FS 4 at the X-point has been detected in $\gamma = 467$~kOe for $T$ = Cu, 574~kOe for $T$ = Au \cite{LaAuSb2_osillation}, and 700~kOe for $T$ = Ag \cite{LaAgSb2 de Hasas-van Alphen, LaAgSb2_osillation2}. Similarly, the close frequency, exhibiting $\alpha = 193$~kOe for $T$ = Cu, $F = 215$~kOe for $T$ = Au \cite{LaAuSb2_osillation}, and $F = 170$~kOe for $T$ = Ag  \cite{LaAgSb2 de Hasas-van Alphen}, suggests that the observed quantum oscillation originates from FS 1 around $\Gamma$ point. The $\beta$-orbit, corresponding to FS 2 Castle-neck-shape close to the R-point, has not been identified in Ag and Au compound. Lastly, dHvA for $T$ = Ag detected more frequencies than that for Au and Cu.

Our investigation through thermodynamic and transport property measurements suggests LaCu$_{x}$Sb$_{2}$ system is an ordinary metal. In particular the transport properties deviate from a typical behavior of Dirac materials. However, the DFT calculations suggest a possibility that LaCu$_{x}$Sb$_{2}$ system may host Dirac fermions, where the contribution of Dirac like charge carriers in transport measurements is probably hidden behind the conventional multiple FS contributions. The exact nature of the topologically nontrivial band structure can be elucidated in further experiments. The angular dependence of quantum oscillations and the measurement under high magnetic field can provide further information about the Fermi surface topology.

\section{Summary and Conclusion}

In summary, we have successfully synthesized LaCu$_{x}$Sb$_{2}$ (0.92 $\le$ $x$ $\le$ 1.12) samples by controlling the initial growth conditions and studied their physical properties by means of electrical resistivity, magnetization, and specific heat. A systematic variation of thermodynamic and transport properties as a function of $x$ clearly demonstrates that physical properties of LaCuSb$_{2}$ are sensitive to the copper stoichiometry. The high quality of the single crystals allows us to measure de Haas-van Alphen (dHvA) and Shubnikov-de Haas (SdH) oscillations. The underlying electronic structures of the ideal copper stoichiometry La$_{1}$Cu$_{1}$Sb$_{2}$ single crystals, inferred from the largest RRR and MR, smallest $\rho_{0}$ value, and highest superconducting transition temperature, are investigated by dHvA and SdH quantum oscillations and DFT calculations. The DFT calculations reveal quasi-2D and 3D Fermi surfaces, in good agreement with results of quantum oscillations, where four frequencies associated with Fermi surfaces are identified. The DFT calculations reveal that bands at the Fermi level are dominated by contributions from the Sb-square net. The La$T$Sb$_{2}$ ($T$ = Cu, Au, and Ag) family hosts a rich variety of properties such as Dirac dispersion, charge density wave, and superconductivity, and the present paper will serve as an important step towards connecting square-net materials and these phenomena. One of great advantages of LaCu$_{x}$Sb$_{2}$ is to provide an opportunity to tune the actual electronic structures via Cu-occupancy, which is supported from the results of quantum oscillations.

\begin{acknowledgments}
This work was supported by the Canada Research Chairs, Natural Sciences and Engineering Research Council of Canada, and Canada Foundation for Innovation program.
\end{acknowledgments}


\begin{thebibliography}{99}


%The Role of Delocalized Chemical Bonding in Square-Net-Based Topological Semimetals
\bibitem{Square-Net-1}
S. Klemenz, S. Lei, and L. M. Schoop, Annu. Rev. Mater. Res. {\bf 49} 185 (2019)

\bibitem{Square-Net-2}
S. Klemenz, A. K. Hay, S. M. L. Teicher, A. Topp, J. Cano, and L. M. Schoop, J. Am. Chem. Soc. {\bf 142} 6350 (2020)

\bibitem{SC and Dirac}
S. J. Ray and L. Alff, Phys. Status Solidi B {\bf 254} 1600163 (2017)

%%%%%%%%%%%%%%%%%%%% Dirac semimetals %%%%%%%%%%%%%%%%%%%%%%%
%Large linear magnetoresistance and magnetothermopower in layered SrZnSb2
\bibitem{Wang2012_SrZnSb2}
K. Wang and C. Petrovic, Appl. Phys. Lett. {\bf 101}, 152102 (2012).

%Magnetic mixed valent semimetal  EuZnSb2 with Dirac states in the band structure
\bibitem{Wang2020_EuZnSb2}
A. Wang, S. Baranets, Y. Liu, X. Tong, E. Stavitski, J. Zhang, Y. Chai, W.-G. Yin, S. Bobev, and C. Petrovic, Phys. Rev. Res. {\bf 2}, 033462 (2020)

%Multiband effects and possible Dirac states in LaAgSbs2
\bibitem{LaAgSb2-multiband}
K. Wang, and C. Petrovic, Phys. Rev. B. {\bf 86} 155213 (2012).

% ref1) Dirac fermions and possible weak antilocalization in LaCuSb2
\bibitem{ref1}
J. R. Chamorro, A. Topp, Y. Fang, M. J. Winiarski, C. R. Ast, M. Krivenkov, A. Varykhalov, B. J. Ramshaw, L. M. Schoop, and T. M. McQueen, APL Mater. {\bf 7}, 121108 (2019)

%Nearly massless Dirac fermions hosted by Sb square net in BaMnSb2
%G-type antiferromagnetic order
\bibitem{Liu2017_BaMnSb2}
J. Liu, J. Hu, H. Cao, Y. Zhu, A. Chuang, D. Graf, D. J. Adams, S. M. A. Radmanesh, L. Spinu, I. Chiorescu, and Z. Mao, Sci. Rep. {\bf 6}, 30525 (2016).

%Quasi-two-dimensional massless Dirac fermions in CaMnSb2
%AFM
\bibitem{He2017_CaMnSb2}
J. B. He, Y. Fu, L. X. Zhao, H. Liang, D. Chen, Y. M. Leng, X. M. Wang, J. Li, S. Zhang, M. Q. Xue, C. H. Li, P. Zhang, Z. A. Ren, and G. F. Chen, Phys. Rev. B {\bf 95}, 045128 (2017).

%Quantum oscillations and coherent interlayer transport in a new topological Dirac semimetal candidate YbMnSb2
%AFM
\bibitem{Wang2018_YbMnSb2}
Y. Y. Wang, S. Xu, L. L. Sun, and T. L. Xia, Phys. Rev. Mater. {\bf 2}, 021201(R) (2018).

%%%%%%%%%% FM and AFM
%A magnetic topological semimetal Sr1−yMn1−zSb2 (y, z < 0.1).
%FM and canted AFM
\bibitem{Liu2017_SrMnSb2}
J. Y. Liu, J.Hu, Q. Zhang, D. Graf, H. B. Cao, S. M. A. Radmanesh, D. J. Adams, Y. L. Zhu, G. F. Cheng, X. Liu, W. A. Phelan, J. Wei, M. Jaime, F. Balakirev, D. A. Tennant, J. F. DiTusa, I. Chiorescu, L. Spinu, and Z. Q. Mao, Nat. Mater. {\bf 16}, 905 (2017).

%Nontrivial Berry phase in magnetic BaMnSb2 semimetal
\bibitem{Huang2017_BaMnSb2}
%AFM along c, FM along ab
S. Huang, J. Kim, W. A. Shelton, E. W. Plummer, and R. Jin, Proc. Natl. Acad. Sci. {\bf 114}, 6256 (2017).

%Observation of a two-dimensional Fermi surface and Dirac dispersion in YbMnSb2
%\bibitem{Kealhofer2018_YbMnSb2}
%AFM implied, not studied
%R. Kealhofer, S. Jang, S. M. Griffin, C. John, K. A. Benavides, S. Doyle, T. Helm, P. J. W. Moll, J. B. Neaton, J. Y. Chan, J. D. Denlinger, and J. G. Analytis, Phys. Rev. B {\bf 97}, 045109 (2018).

%Nontrivial topology in the layered Dirac nodal-line semimetal candidate SrZnSb2 with distorted Sb square nets 
%non-magnetic
\bibitem{Liu2019_SrZnSb2}
J. Liu, P. Liu, K. Gordon, E. Emmanouilidou, J. Xing, D. Graf, B. C. Chakoumakos, Y. Wu, H. Cao, D. Dessau, Q. Liu, and N. Ni, Phys. Rev. B {\bf 100}, 195123 (2019).

%Canted antiferromagnetic phases in the candidate layered Weyl material EuMnSb2
%AFM
\bibitem{Wilde2022_EuMnSb2}
J. M. Wilde, S. X. M. Riberolles, A. Das, Y. Liu, T. W. Heitmann, X. Wang, W. E. Straszheim, S. L. Bud'ko, P. C. Canfield, A. Kreyssig, R. J. McQueeney, D. H. Ryan, and B. G. Ueland, Phys. Rev. B {\bf 106}, 024420 (2022).

%%%%%%%%%%%%% CDW %LaAgSb2-Myers1999, LaAgSb2-C.song2003, AuCusub-LaAgSb2 %Systematic study of anisotropic transport and magnetic properties of RAgSb2 (R=Y, La?Nd, Sm, Gd?Tm)
\bibitem{LaAgSb2-Myers1999}
K. D. Myers, S. L. Bud'ko, I. R. Fisher, Z. Islam, H. Kleinke, A. H. Lacerda and P. C. Canfield, J. Magn. Magn. Mater. {\bf 205}, 27 (1999).

%Charge-density-wave orderings in LaAgSb2: An x-ray scattering study
\bibitem{LaAgSb2-C.song2003}
C. Song, J. Park, J. Koo, K-B Lee, J. Y. Rhee, S. L. Bud'ko, P. C. Canfield, B. N. Harmon, and A. I. Goldman, Phys. Rev. B {\bf 68}, 035113 (2003). 

%Weak charge-density-wave transition in LaAgSb2 investigated by transport, thermal, and NMR studies
\bibitem{LaAgSb2-C.S.Lue2007} 
C. S. Lue, Y. F. Tao, K. M. Sivakumar, and Y. K. Kuo, J. Phys.: Condens. Matter {\bf 19} 406230 (2007). 

%Revealing Extremely Low Energy Amplitude Modes in the Charge-Density-Wave Compound LaAgSb2
\bibitem{LaAgSb2-R.Y.Chen2017}
R. Y. Chen, S. J. Zhang, M. Y. Zhang, T. Dong, and N. L. Wang, Phys. Rev. Lett. {\bf 118}, 107402 (2017).

%Successive destruction of charge density wave states by pressure in LaAgSb2
\bibitem{LaAgSb2_osillation2}
K. Akiba, H. Nishimori, N. Umeshita, and T. C. Kobayashi, Phys. Rev. B {\bf 103}, 085134 (2021).

%Characterization of the charge density wave transition and observation of the amplitude mode in LaAuSb2
\bibitem{LaAuSb2-2019} 
C. N. Kuo, D. Shen, B. S. Li, N. N. Quyen, W. Y. Tzeng, C. W. Luo, L. M. Wang, Y. K. Kuo, and C. S. Lue, Phys. Rev. B. {\bf 99} 235121 (2019).

%Tuning of charge density wave transitions in LaAuxSb2 by pressure and Au stoichiometry
\bibitem{LaAuSb2-2020}
Li Xiang, D. H. Ryan, W. E. Straszheim, P. C. Canfield, and S. L. Bud’ko, Phys. Rev. B. {\bf 102}, 125110 (2020).

%22Pressure-dependent modifications in the LaAuSb2 charge density wave system
\bibitem{LaAuSb2-2021}
G. Lingannan, B. Joseph, P. Vajeeston, C. N. Kuo, C. S. Lue, G. Kalaiselvan, P Rajak, and S Arumugam, Phys. Rev. B. {\bf 103} 195126 (2021).

%Coexistence of Dirac fermion and charge density wave in the square-net-based semimetal LaAuSb2
\bibitem{LaAuSb2_osillation}
X. Wu, Z. Hu, D. Graf, Y. Liu, C. Deng, H. Fu, A. K. Kundu, T. Valla, C. Petrovic, and A. Wang, Phys. Rev. B {\bf 108}, 245156 (2023).

%%%%%%%%%%%%% SC
\bibitem{LaAgSb2.K.Akia}
K. Akiba, N. Umeshita, and T.C. Kobayashi, Phys. Rev. B {\bf 106}, L161113 (2022).

%Interplay between charge density wave order and superconductivity in LaAuSb2 under pressure
\bibitem{LaAuSb2-Tc}
F. Du, H. Su, S. S. Luo, B. Shen, Z. Y. Nie, L. C. Yin, Y. Chen, R. Li, M. Smidman, and H. Q. Yuan, Phys. Rev. B. {\bf 102}, 144510 (2020).


%LaCuSb2
%ref2 )Phonon-mediated superconductivity in the Sb square-net compound LaCuSb2
%28
\bibitem{ref2}
K. Akiba, and T. C. Kobayashi, Phys. Rev. B. {\bf 107}, 245117 (2023).

%ref3) Fragile superconductivity in a Dirac metal
%29
\bibitem{ref3}
C. J. Lygouras, J. Zhang, J. Gautreau, M. Pula, S. Sharma, S. Gao, T. Berry, T. Halloran, P. Orban, G. Grissonnanche, J. R. Chamorro, K. Mikuri, D. K. Bhoi, M. A. Siegler, K. K. Livi, Y. Uwatoko, S. Nakatsuji, B. J. Ramshaw, Y. Li, G. M. Luke, C. L. Broholm, and T. M. McQueen, arXiv:2307.01976. 

%ref7) Magnetic and transport properties of dense Kondo systems, CeTSb 2 (T=Ni, Cu, Pd and Ag) \
%19
\bibitem{ref7}
Y. Muro, N. Takeda, and M. Ishikawa, J Alloy. Comp. {\bf 257}, 23 (1997).  

% Effect of Au or Cu susbtitution.
%Chemical substitution effect on CDW state in LaAgSb2
%16
\bibitem{AuCusub-LaAgSb2}
S. Masubuchi, Y. Ishii, K. Ooiwa, T. Fukuhara, F. Shimizu, and H .Sato, JPS Conf. Proc. {\bf 3} 011053 (2014).


%Effects of Physcial and chemical Pressure on Charge Density Wave Transitions in LaAg1-xAuxSb2 Single Crystals
%22
\bibitem{LaAg1-xAuxSb2-2022}
L. Xiang, D. H. Ryan, P. C. Canfield, and S. L. Bud’ko, crystals {\bf 12} 1693 (2022).


%LaAgSb2 about band structure
%de Haas–van Alphen and Shubnikov–de Haas oscillations in RAgSb2 (R=Y, La-Nd, Sm)
\bibitem{LaAgSb2 de Hasas-van Alphen}
K. D. Myers, S. L. Bud'ko, V. P. Antropov, B. N. Harmon, P. C. Canfield, and A. H. Lacerda, Phys. Rev. B. {\bf 60}, 13371 (1999).


%Electronic Band Structure and Surface States in Dirac Semimetal LaAgSb2
\bibitem{LaAgSb2-2022}
M. Rosmus, N. Olszowska, Z. Bukowski, P. Starowicz, P. Piekarz, and A. Ptok, Materials {\bf 15}, 7168 (2022).


%Stoichiometry of LaAuSb2

%Ternary antimonides LnM1-xSb2 with Ln=La-Nd, Sm, Gd, Tb and M=Mn, Co, Au, Zn, Cd
%30
\bibitem{LaAuSb2-1996}
P. Wollesen, W. Jeitschko, M. Brylak, and L. Dietrich, Jour. All. Comp. {\bf 245} L5-L8 (1996).



%Electronic band calculation of LaTSb2 (T=Cu,Ag,Au)
%13
\bibitem{LaTSb2-2014}
I. Hase and T. Yanagisawa, Phys. Procedia. {\bf 58}, 42 (2014).


%Computational methods, DFT

%QUANTUM ESPRESSO: A modular and open-source software project for quantum simulations of materials
%30
\bibitem{Giannozzi2009}
P. Giannozzi, S. Baroni, N. Bonini, M. Calandra, R. Car, C.
Cavazzoni, D. Ceresoli, G. L. Chiarotti, M. Cococcioni, I. Dabo
et al., J. Phys.: Condens. Matter {\bf 21}, 395502 (2009).


%Generalized Gradient Approximation Made Simple 
%31
\bibitem{JPPerdew1996_PRL}
J. P. Perdew, K. Burke, M. Ernzerhof, Phys. Rev. Lett. {\bf 77}, 3865 (1996).

%Optimized norm-conserving Vanderbilt pseudopotentials,
%32
\bibitem{Hamann_PRB2013}
D. R. Hamann, Phys. Rev. B {\bf 88}, 085117 (2013).


%Special points for Brillouinzone integrations
%33
\bibitem{HJMonkhorst1976}
H. J. Monkhorst and J. D. Pack, Phys. Rev. B {\bf 13}, 5188 (1976).

%Thermal Contraction and Disordering of the Al(110) Surface
%34
\bibitem{Marzari99}
N. Marzari, D. Vanderbilt, A. De Vita, and M. C. Payne, Phys. Rev. Lett. {\bf 82}, 3296 (1999).

%FermiSurfer: Fermi-surface viewer providing multiple representation schemes
%35
\bibitem{Kawamura2019_FS}
M. Kawamura, Comput. Phys. Commun. {\bf 239}, 197 (2019).


%ref4) Thermoelectric Properties of Ternary Rare-Earth Copper Antimonides LaCuxSb2 (0:9 < x < 1.3)_ occupancy controlled by initial concentration. arc-melt
%36
\bibitem{ref4}
M. Ohta, Materials Transactions, {\bf 50}, 7, 1881 (2009).

%ref5) RCu1+xSb2  (R = La, Ce, Pr, Nd, Sm, Gd, Tb, Dy, Ho and Y ) phases with defect CaBe2Ge2 - type structure 
\bibitem{ref5}
X. X. Yang, Y. M. Lu, S. K. Zhou, S. Y. Mao, J. X. Mi, Z. Y. Man and J. T. Zhao, Materials Science Forum {\bf 475-479} 861 (2005). 

%ref6) On the crystal structure and magnetic properties of the ternary rare earth compounds RETSb2 with RE-rare earth and T-Ni, Pd, Cu and Au_poly 
\bibitem{ref6}
O. Sologub, K. Hiebl and P. Rogl, H. Noel, O. Bodak, J. Alloy. Comp. {\bf 210}, 153 (1994). 


%ref8)  Besetzungsvarianten der CaMnBi2-Struktur: Zur Kenntnis der Verbindungen LaZn0,52Sb2, LaCo0,68Sb2, LaMn*Sb2 (0,65 0,76) und LaCu^-S^ (0,82 0,87)
\bibitem{ref8}
G. Cordier, H. Schäfer and P. Woll, Z. Naturforsch, {\bf 40b}, 1097 (1985).


%Dirac-like band structure of LaTESb2 (TE = Ni, Cu, and Pd) superconductors by DFT calculations
\bibitem{LaTESb2_Ruszala2018}
P. Rusza{\l}a, M. J. Winiarski, M. Samsel-Czeka{\l}a, Comput. Mater. Sci. {\bf 154}, 106 (2018).

%Dirac-like electronic-band dispersion of LaSb2 superconductor and its counterpart LaAgSb2
\bibitem{LaAgSb2_Ruszala2018}
P. Rusza{\l}a, M. J. Winiarski, M. Samsel-Czeka{\l}a, Acta Phys. Pol. A {\bf 138}, 748 (2020).

\end{thebibliography}
\end{document}